\documentclass[11pt,a4paper]{article}
\usepackage{jheppub}
\usepackage{amssymb,amsmath,amsfonts}
\usepackage{epsfig}

\def\clock{{\count0=\time
           \divide\count0 60
           \ifnum\count0<10 0\fi\the\count0
           \multiply\count0 -60 \advance\count0 \time
           :\ifnum\count0<10 0\fi \the\count0
         }}
\newcommand{\timestamp}{{\small\vbox{\hbox{\tt\jobname.tex}
\hbox{\the\day/\the\month/\the\year, \clock}}}}

\newcommand{\CB}{\mathcal{B}}
\newcommand{\CC}{\mathcal{C}}

\newcommand{\CH}{\mathcal{H}}

\newcommand{\CL}{\mathcal{L}}
\newcommand{\CM}{\mathcal{M}}
\newcommand{\CN}{\mathcal{N}}
\newcommand{\CO}{\mathcal{O}}
\newcommand{\CP}{\mathcal{P}}

\newcommand{\CS}{\mathcal{S}}

\newcommand{\Z}{\mathbb{Z}}
\newcommand{\C}{\mathbb{C}}
\newcommand{\R}{\mathbb{R}}

\newcommand{\nn}{\nonumber}

\newcommand{\spa}{\ , \ \ }

\newcommand{\ds}{\displaystyle}

\newcommand{\matrto}[4]{\left( \begin{array}{cc} #1 & #2 \\
#3 & #4 \end{array} \right) }

\newcommand{\ads}{\mbox{AdS}}
\newcommand{\cft}{\mbox{CFT}}
\newcommand{\sphere}{\mbox{S}}

\newcommand{\be}{\begin{eqnarray}}
\newcommand{\ee}{\end{eqnarray}}

\newcommand{\grp}[1]{\mathrm{#1}}

\newcommand{\grSU}{\grp{SU}}
\newcommand{\grU}{\grp{U}}

\usepackage{color}

\usepackage{hyperref}

\title{Finite-size corrections for quantum strings on \boldmath $\ads_4\times \C P^3$}
\author[a]{Davide Astolfi,}
\author[b]{Valentina Giangreco M. Puletti,}
\author[a]{Gianluca Grignani,}
\author[b]{Troels Harmark,}
\author[c]{and Marta Orselli}

\affiliation[a] {Dipartimento di Fisica, Universit\`a di Perugia,\\
I.N.F.N. Sezione di Perugia,\\
Via Pascoli, I-06123 Perugia, Italy}
\affiliation[b]  {NORDITA\\
Roslagstullsbacken 23,
SE-106 91 Stockholm, Sweden}
\affiliation[c]{The Niels Bohr Institute  \\
\sl  Blegdamsvej 17, 2100 Copenhagen \O , Denmark }

\emailAdd{astolfi@pg.infn.it}
\emailAdd{valentina@nordita.org}
\emailAdd{grignani@pg.infn.it}
\emailAdd{harmark@nordita.org}
\emailAdd{orselli@nbi.dk}

\abstract{
We revisit the calculation of curvature corrections to the pp-wave energy of type IIA string states on $\ads_4\times \C P^3$ initiated in arXiv:0807.1527.
Using the near pp-wave Hamiltonian found in arXiv:0912.2257, we compute the first non-vanishing correction to the energy of a set of bosonic string states at order $1/R^2$, where $R$ is the curvature radius of the background.  The leading curvature corrections give rise to cubic, order $1/R$, and quartic, order $1/R^2$,
terms in the Hamiltonian, for which we implement the appropriate normal ordering prescription. Including the contributions from all possible fermionic and bosonic string states, we find that there exist
logarithmic divergences in the sums over mode numbers which cancel between
the cubic and quartic Hamiltonian. We show that from the form of the cubic Hamiltonian it is natural to require that the cutoff for summing over heavy modes must be twice the one for light modes. With this prescription the strong-weak coupling interpolating function $h(\lambda)$, entering the magnon dispersion relation, does not receive a one-loop correction, in agreement with the algebraic curve spectrum. However, the single magnon dispersion relation exhibits finite-size exponential corrections.}


\begin{document}
\maketitle
\flushbottom

\setcounter{page}{1}


\section{Introduction and summary}
\label{sec:intro}

The AdS/CFT correspondence~\cite{Maldacena:1997re,Gubser:2002tv,Witten:1998qj} predicts that the energies of excited states of superstrings living on certain backgrounds should match the anomalous dimensions of operators of the dual gauge field theory.
In the planar limit of the correspondence the string coupling is zero, but the string still lives on a non-trivial curved background which means that the corresponding two-dimensional world-sheet theory is not free. In general this makes computing the superstring spectrum in such backgrounds a difficult problem.

However, the complicated interactions in the superstring world-sheet theory vanish when taking a Penrose limit of the geometry where the string lives~\cite{Metsaev:2001bj,Berenstein:2002jq}. It is then possible to compute curvature corrections to the free string spectrum as a perturbative expansion in inverse powers of the curvature radius $R$ of the background.

This approach was developed by Callan et al. in~\cite{Callan:2003xr,Callan:2004uv} for the $\ads_5/\cft_4$ duality, so far the most well-understood example of a string/gauge duality, which states the correspondence between type IIB superstring theory on $\ads_5 \times \sphere^5$ and $\mathcal N=4$ super Yang-Mills (SYM) theory in four dimensions. The results of~\cite{Callan:2003xr,Callan:2004uv} and the corresponding analysis on the gauge theory side \cite{Beisert:2003tq},
have been fundamental in the study of the integrability of the AdS/CFT correspondence. Refs.~\cite{Callan:2003xr,Callan:2004uv} produced the first evidence of the famous three loop discrepancy between anomalous dimensions  of gauge theory operators and energies of the dual string states. This discrepancy was subsequently understood and solved by the inclusion of the dressing factor, that interpolates between weak and strong coupling, in the Bethe equations that describe the spectrum of the gauge and the string theory~\cite{Minahan:2002ve, Arutyunov:2004vx, Hernandez:2006tk, Beisert:2006ez}.

More recently a new exact duality between gauge and string theory has been proposed, type IIA superstring on $\ads_4 \times \C P^3$
is dual to a certain limit of ABJM-theory~\cite{Aharony:2008ug}.
ABJM theory is an $\mathcal N=6$ Chern- Simons-matter gauge theory  dual to M-theory compactified on $\ads_4 \times \sphere^7/\Z_k$. It has a $\grU(N) \times \grU (N )$ gauge symmetry with Chern-Simons like kinetic terms at level $k$ and $-k$ and in the region where the 't Hooft coupling $\lambda=\frac{N}{k}$ is  $1\ll \lambda\ll k^4$, the gravity side can be effectively described as a type IIA superstring on $\ads_4 \times \C P^3$.

Having this new $\ads_4/\cft_3$ duality naturally brings up the question of its integrability which in fact has received a lot of attention~\cite{Minahan:2008hf,Gaiotto:2008cg,Grignani:2008is,Grignani:2008te,Astolfi:2008ji,McLoughlin:2008he,
Sundin:2008vt,Zarembo:2009au,Minahan:2009aq,
Minahan:2009wg}, \cite{Arutyunov:2008if,Stefanski:2008ik,Bak:2008cp,Kristjansen:2008ib,Zwiebel:2009vb, Minahan:2009te,Bak:2009tq,Sorokin:2010wn,Sorokin:2011mj}.
In particular an all-loop asymptotic Bethe ansatz has been proposed~\cite{Gromov:2008qe} and recently a set of functional equations in the form of a Y-system has been formulated also for the $\ads_4/\cft_3$ duality~\cite{Gromov:2009tv,Bombardelli:2009ns,Gromov:2009bc,Bombardelli:2009xz}.

In this Paper we fill a gap in the study of the integrability of this theory, by performing a complete calculation, as that in \cite{Callan:2003xr,Callan:2004uv} for the $\ads_5/\cft_4$ duality, of the curvature corrections to the pp-wave energy of a set of bosonic type IIA string states on $\ads_4\times \C P^3$. This investigation was initiated in \cite{Astolfi:2008ji} and subsequently revisited in \cite{Sundin:2008vt,Sundin:2009zu}. However, only in \cite{Astolfi:2009qh} was the interacting Hamiltonian for quantum strings in a near plane wave limit of $\ads_4\times \C P^3$ computed in full, including all the fermionic contributions. This is given as
a perturbative expansion in terms of inverse powers of the curvature radius
\footnote{In our notation $R$ is the radius of $\C P^3$, cf. appendix \ref{app:geometrical_setup}.}
$R$  of the form
\be
\label{eq0}
H &=& H_2+ {1\over R} H_3 +{1\over R^2} H_4+\mathcal{O}(1/R^3)
\cr
& =&  H_{2,B}+H_{2,F} +  {1\over R} \left( H_{3,B}+ H_{3,BF}\right) + {1\over R^2} \left( H_{4,B}+ H_{4,F}+ H_{4,BF}\right) +\mathcal{O}(1/R^3)
\ee
The quadratic Hamiltonian $H_2$ is nothing but the plane-wave free Hamiltonian where the fermionic and bosonic fields are completely decoupled~\cite{Gaiotto:2008cg, Nishioka:2008gz, Grignani:2008is,Grignani:2009ny}. A characteristic of this theory is that at the pp-wave level, the 8 massive bosons and 8 massive fermions have different world sheet masses. 4 bosons and 4 fermions are ``heavy", whereas the remaining 4 bosons and 4 fermions are ``light" with a world-sheet mass which is half of that of the heavy modes.

At the next-to-leading order, the computation involves a new feature compared to the case of type IIB string theory on $\ads_5\times S^5$, namely, a $1/R$ correction cubic in the number of fields, appears in the Hamiltonian~\cite{Astolfi:2008ji}. $H_{3,B}$ has three bosonic fields and $H_{3,BF}$ one bosonic and two fermionic fields. The perturbed energy of the string states is computed through standard perturbation theory. The $1/R$ correction manifests itself only at the second order in the perturbative expansion, order $1/R^2$, since at the first order it gives a vanishing contribution.
Finally the order $1/R^2$ term in \eqref{eq0} is an interaction quartic in the number of fields contributing to the energies at first order. $H_{4,B}$ has four bosonic fields, $H_{4,F}$ four fermionic fields  and $H_{4,BF}$ two bosonic and two fermionic fields.
The first non-trivial correction to the energy of a string state then appears at order $1/R^2$ and reads
\be
\label{eq1}
E_{e}^{(2)}=\frac{1}{R^2}\left(\sum_{|i\rangle}\frac{\left|\langle i|H_{3}|e \rangle\right|^2}{E^{(0)}_{|e \rangle}-E^{(0)}_{|i\rangle}}+\langle e |H_{4}| e\rangle\right)
\ee
where $|e\rangle$ is a certain external state with zeroth order energy $E^{(0)}_{|e\rangle}$ and $|i\rangle$ is an intermediate state with zeroth order energy $E^{(0)}_{|i\rangle}$. Eq. \eqref{eq1}
corresponds, on the dual gauge theory side, to finite size corrections at order $1/J$
to the anomalous dimension of gauge theory operators (where $J$ corresponds to the $R$-charge of these operators).
The near-plane wave configuration around which we are perturbing is the one described in \cite{Grignani:2008is}, where the string is a point-like configuration sitting on the equator of one of the two 2-spheres  embedded in $\C P^3$ and fast rotating with an angular momentum $J$.

The first term in \eqref{eq1} is a completely new feature of the $\ads_4/\cft_3$ duality, it can be considered as a one-loop effect generated by the cubic Hamiltonian. It gives
logarithmic divergences in the sums over intermediate states. These divergences must be canceled  by the second term in \eqref{eq1}, generated by the quartic Hamiltonian, which, consequently, cannot be normal ordered, as it is for type IIB superstring on $\ads_5 \times \sphere^5$~\cite{Callan:2003xr}.
The starting point for the computation of the perturbed spectrum \eqref{eq1} is the Hamiltonian obtained in~\cite{Astolfi:2009qh}, which we now need to quantize and diagonalize.
The key observation is that such an Hamiltonian, as well as the Lagrangian, have been derived as classical objects. This implies that when one promotes the classical fields to quantum operators, an ordering ambiguity appears. Such an ambiguity in the normal-ordering prescription was already discussed in~\cite{Callan:2003xr}  for the $\ads_5 \times \sphere^5$ string case. There it was shown, using arguments based on the fact that the $\ads_5 \times \sphere^5$ is a maximally supersymmetric background, that the quartic Hamiltonian is normal ordered (the sextic Hamiltonian is most likely not).
In the $\ads_4\times\C P^3$ case however, we will show that, in order to obtain a finite spectrum for string states, one has to introduce a non-trivial normal ordering prescription for the quantum Hamiltonian. The appropriate normal ordering prescription turns out to be the Weyl prescription, the completely symmetric (or antisymmetric in the case of fermions) prescription. With such a prescription, when all contributions are assembled, divergences cancel, leaving a finite result which we compute in full.

$E_{e}^{(2)}$ will be computed for two-oscillator states in the $\grSU(2)\times \grSU(2)$ sector of $\C P^3$, as well as for a state with a generic number of oscillators in the $\grSU(2)\times \grSU(2)$ sector where the mode numbers are all different. Moreover we will
consider two-oscillator bosonic states outside the $\grSU(2)\times \grSU(2)$ sector, but still inside $\C P^3$.

We denote the two-oscillator states in the $\grSU(2)\times \grSU(2)$ sector as the state $|s\rangle$ where the two oscillators are in the same $\grSU(2)$ sector and the state $|t\rangle$ where there is one oscillator in each $\grSU(2)$ sector~\cite{Astolfi:2008ji}.
Our results for these two states are
\begin{equation} \label{senergyintro}
E_s^{(2)}=-\frac{8\,n^2\left[\left(\omega_n-\frac{c}{2}\right)^2-\frac{c^2}{2}\right]}{R^2 c^3\omega_n^2}-\frac{8 n^2}{R^2 c\, \omega_n}\sum_{q=1}^\infty\left[1-(-1)^q\right]K_0(\pi c q)
 \end{equation}
and
\begin{equation} \label{tenergyintro}
E_t^{(2)}=-\frac{8\,n^2\left(\omega_n-\frac{c}{2}\right)^2}{R^2 c^3\omega_n^2}-\frac{8 n^2}{R^2 c\, \omega_n}\sum_{q=1}^\infty\left[1-(-1)^q\right]K_0(\pi c q)
\end{equation}
where $\omega_n=\sqrt{n^2+\frac{c^2}{4}}$ is the pp-wave energy of a light mode, $c={4J\over R^2}= {J\over \pi\sqrt{2\lambda}}\equiv {1\over\pi \sqrt{2\lambda'}}$ and $K_0(x)$ is the modified Bessel function of the second kind.
The first terms in both Eqs.~\eqref{senergyintro} and \eqref{tenergyintro} were already computed in~\cite{Astolfi:2008ji} where only the purely bosonic part in the $\grSU(2)\times \grSU(2)$ sector of the full Hamiltonian was considered and the divergences appearing in the computation were treated using zeta function regularization and assuming normal ordering for the quartic Hamiltonian. Here instead we perform the computation including in the Hamiltonian all the bosonic and fermionic fields, as derived in~\cite{Astolfi:2009qh}. We show that, after using the appropriate normal ordering prescription, we obtain the result given in Eqs.~\eqref{senergyintro} and \eqref{tenergyintro}, where the last sums containing the Bessel functions are exponentially suppressed in the limit of large $c$ (small $\lambda'=\lambda/J^2$).

The results \eqref{senergyintro} and \eqref{tenergyintro} are free of divergences, but there is at the moment an interesting discussion among the scientific community about how to correctly regularize the sums over mode numbers~\cite{Krishnan:2008zs, McLoughlin:2008ms, Alday:2008ut,  Gromov:2008fy, Bandres:2009kw} leading to one-loop energies as those in \eqref{senergyintro} and \eqref{tenergyintro}. We face here the same issues, however, from our quantum string world-sheet calculation we have evidence that the most natural prescription is the one proposed in~\cite{Gromov:2008qe,Gromov:2008fy, Bandres:2009kw} which is in agreement with the algebraic curve spectrum and the results inferred from gauge theory calculations~\cite{Gromov:2008bz,Leoni:2010tb}. We shall give details of our regularization prescription, which leads to the energies \eqref{senergyintro} and \eqref{tenergyintro} for the $|s\rangle$ and $|t\rangle$ states, in Section \ref{sec:string_finitesize_en}.

After fixing the light-cone gauge and $\kappa$-symmetry, the residual symmetries of the theory fix the form of the light-magnon dispersion relation
\be
\label{magnon_disp_rel}
E= \sqrt{{1\over 4} +4 h^2(\lambda) \sin^2 {p\over 2}}
\ee
 but not the function $h(\lambda)$~\cite{Gaiotto:2008cg, Grignani:2008is, Nishioka:2008gz}, which interpolates between strong and weak gauge theory coupling regime.
In the $\ads_5/\cft_4$ duality the magnon dispersion relation is $E=\sqrt{1 +f(\lambda) \sin^2 {p\over 2}}$~\cite{Beisert:2004hm,Beisert:2005tm}, where the function $f(\lambda)$ turns out to be equal to $\frac{\lambda}{\pi^2}$ both at strong coupling and weak coupling. In contrast, earlier studies of the BMN limit in the $\ads_4/\cft_3$ duality revealed that the function $h(\lambda)$ behaves as $\lambda +\mathcal O(\lambda^4)$ at weak coupling~\cite{Minahan:2008hf, Gaiotto:2008cg, Nishioka:2008gz} and as $\sqrt{{\lambda\over 2}}+\mathcal O(\lambda^0)$ at strong coupling~\cite{Gaiotto:2008cg, Grignani:2008te, Nishioka:2008gz}.
 Furthermore, the interpolating function has been computed up to four loops on the gauge theory side in~\cite{Minahan:2009aq, Minahan:2009wg, Leoni:2010tb}.
 The semiclassical analysis of the folded and spinning strings has led the authors of~\cite{McLoughlin:2008ms, Alday:2008ut, Krishnan:2008zs, McLoughlin:2008he} to conclude that there should be a one-loop correction of order $\lambda^0$ entering in $h(\lambda)$ at strong coupling, namely that
 \be
 h(\lambda) = \sqrt{{\lambda\over 2}} +a^{\rm WS}_1 +\mathcal O \left({1\over \sqrt{\lambda}}\right) \qquad \text{with} ~~~ a^{\rm WS}_1=- {\log{2} \over 2\pi} \,, \qquad \lambda\gg1
 \ee
where the superscript $\rm WS$ stands for world-sheet.

Using instead predictions based on the algebraic curve, the Bethe ansatz of~\cite{Gromov:2008qe, Gromov:2008fy} and the extrapolation to strong coupling of the all loop ansatz of~\cite{Leoni:2010tb}, it seems that the first non-trivial contribution to the interpolating function should start at higher order, namely that
 \be
h(\lambda) = \sqrt{{\lambda\over 2}} +a^{\rm AC}_1 +\mathcal O \left({1\over \sqrt{\lambda}}\right) \qquad \text{with} ~~~ a^{\rm AC}_1=0\,, \qquad \lambda\gg1
 \ee
 where the superscript $\rm AC$ stands for algebraic curve.

The two different values for the one-loop correction $a_1$ to $h(\lambda)$ originate from different regularization schemes employed in the two types of analysis. In particular two different prescriptions have been proposed for summing over mode numbers~\cite{Gromov:2008fy, Bandres:2009kw}. The important point is that the two prescriptions differ by a constant factor, which affects the result and, in turns, is experienced as a one-loop contribution, {\it i.e.} $a^{\rm WS}_1\,,  a^{\rm AC}_1 \sim \mathcal{O} (\lambda^0)$, to the interpolating function $h(\lambda)$, thus affecting the magnon dispersion relation \eqref{magnon_disp_rel}. The motivation behind the different
prescriptions is essentially that in the one giving $a^{\rm WS}_1$ one treats all modes on an equal footing, while in the one giving  $a^{\rm AC}_1$ one distinguishes between light and heavy  excitations of the theory. The idea behind the latter regularization is that heavy excitations are not fundamental but rather bound
states of two light fundamental modes~\cite{Zarembo:2009au}. This leads to the choice of a different cutoff for
the two kinds of excitations. Here we show that the cubic interaction Hamiltonian naturally leads to the cutoff
on heavy modes to be twice that on light modes.

The curvature corrections to the string state energies we compute in this Paper, would feel the presence of an $a_1$ term in $h(\lambda)$. In fact in the BMN limit the momentum in the dispersion relation \eqref{magnon_disp_rel} is $p=\frac{2\pi n}{J}$ and expanding for large $J$ using a non vanishing $a_1$ in $h(\lambda)$
  gives
\begin{eqnarray}\label{dispersioncorrected}
E&=& \sqrt{{1\over 4} + 4\left(\sqrt{\frac{\lambda}{2}}+a_1\right)^2\sin^2 {p\over 2}}\cr&\simeq&\sqrt{{1\over 4} + 2\lambda' n^2\pi^2}+ \frac{\sqrt{2}a_1}{J} \left[4 \pi ^2  n^2
   \sqrt{\lambda'}-16 \pi ^4
  n^4 {\lambda'}^{3/2}+96 \pi ^6  n^6 {\lambda'}^{5/2}+\mathcal{O}\left({\lambda'}^{7/2}\right)\right]\cr&&
\end{eqnarray}
namely a $\frac{1}{J}= \frac{4}{c R^2}$ term with half integer powers of $\lambda'$. Such a term could in principle appear in \eqref{senergyintro} and \eqref{tenergyintro}, which are the finite size energies of two light magnons, but with our regularization it does not. The expansion in powers of $\lambda'$ of the first terms in \eqref{senergyintro} and \eqref{tenergyintro} gives integer powers of $\lambda'$ while the sum with the Bessel function gives non analytic terms that go like  $\sim e^{-{J\over \sqrt{\lambda}}}$. We have thus to conclude that $a_1=0$, in agreement with the algebraic curve calculation.

The sum in \eqref{senergyintro} and \eqref{tenergyintro} appears diagonally also in the corrections to the energy of states with a generic number of oscillators with different level number and also in the finite size corrections to the energy of states outside the $\grSU(2)\times \grSU(2)$ sector, see Sections~\ref{sec:string_finitesize_en} and \ref{sec:outside_SU2xSU2}. Therefore, it appears in all the string states we have considered and can be ascribed to a finite size correction of the single magnon dispersion relation~\eqref{magnon_disp_rel}. Actually further evidence of this interpretation is provided  by the fact that if one considers a single oscillator state in one of the $\grSU(2)$'s,  relaxing the level matching condition, in its spectrum the same sum as in \eqref{senergyintro} and \eqref{tenergyintro} appears, see Section~\ref{sec:string_finitesize_en}. This one-loop effect should be generated, in the dual gauge theory, by wrapping interactions  and should in pri!
 nciple be computable applying  L\"uscher's corrections to the study of the worldsheet QFT of the superstring (see the recent review on the subject~\cite{Janik:2010kd} and reference therein). This is then another instance of the exponentially suppressed finite size corrections to the magnon dispersion relation that, for type IIA superstring on $\ads_4 \times \C P^3$, were first computed in the giant magnon limit in \cite{Grignani:2008te} (see also \cite{Abbott:2008qd,Abbott:2009um}) and derived from L\"uscher's corrections in \cite{Bombardelli:2008qd,Lukowski:2008eq,Ahn:2008wd,Ahn:2010eg}. In conclusion, our results are compatible with a single magnon dispersion relation which in the near BMN limit has the form
\begin{eqnarray}
\label{magnon_disp_rel_corrected}
E&=& \sqrt{{1\over 4} +4 \left(\sqrt{\frac{\lambda}{2}}+\CO\left(\frac{1}{\sqrt{\lambda}}\right)\right)^2 \frac{n^2\pi^2}{J^2}}-\frac{\pi n^2\sqrt{2\lambda'} }{J\, \sqrt{{1\over 4} + 2\lambda' n^2\pi^2}}\sum_{q=1}^\infty\left[1-(-1)^q\right]K_0\left(\frac{q}{\sqrt{2\lambda'}}\right)\cr&&
\end{eqnarray}

The Paper is organized as follows.
In Section~\ref{spoint} we describe the procedure, the light-cone gauge fixing and derive the mode expansion and the pp-wave spectrum.

In Section~\ref{sec:string_finitesize_en} we explicitly compute the energies of the string states $|s\rangle$ and $|t\rangle$ deriving  the equations \eqref{senergyintro} and \eqref{tenergyintro} and those for a generic number of oscillators in the $\grSU(2)\times \grSU(2)$ sector when all the mode numbers are different.
In Section~\ref{regularization} we describe the regularization procedure we use, by showing that the cubic Hamiltonian requires that the cutoff for summing over heavy modes must be twice that for light modes. We also comment on other regularization prescriptions which would lead to \eqref{dispersioncorrected} with a non vanishing $a_1$.

In Section~\ref{sec:no} we discuss our normal ordering prescription, showing that it is consistent with the pp-wave algebra of generators.
In Section~\ref{sec:outside_SU2xSU2} we compute the first finite size correction also for states outside the $\grSU(2)\times \grSU(2)$ sector of $\C P^3$ showing that they are consistent and finite.
In Section~\ref{sec:conclusions} we draw our conclusions.

The appendices are devoted to the review of the geometrical setup, the Penrose limit we use, the explicit expressions of the interacting Hamiltonian used in the calculations, to the conventions on gamma matrices and to the small $c$ expansion of the sum in \eqref{senergyintro}.


\section{Preliminaries}
\label{spoint}

The type IIA superstring we are interested in, lives on the $\ads_4\times \C P^3$ background
with a two-form and four-form Ramond-Ramond flux turned on. The metric and the explicit expressions for the field strengths are given in the appendix \ref{app:geometrical_setup}.
As explained in the introduction, we want to compute near-plane-wave corrections to the energy of such a string, with a particular interest for states which are in the $\grSU(2)\times \grSU(2)$ sector of $\C P^3$. These corrections correspond to finite-size corrections to the anomalous dimension of the dual gauge theory operators and are quantum corrections since they are computed by means of quantum mechanics perturbation theory. $1/J$  corrections to the anomalous dimension of gauge theory operators correspond to the near-plane wave corrections to the spectrum of string states up to order $1/R^2=1/(4\pi J \sqrt{2 \lambda'})$, where $R$ is the $\C P^3$ radius and $\lambda'=\lambda/J^2$ is kept fixed.

Hence we are working with a perturbative, i.e. large $R$, analysis around a plane-wave configuration which is represented by a point-like type IIA string moving in the $\grSU(2)\times\grSU(2)$ subsector of $\C P^3$ and along the time direction $\R_t$ on $\ads_4$~\cite{Nishioka:2008gz, Gaiotto:2008cg, Grignani:2008is}. The specific plane-wave background~\footnote{An extensive study of all possible pp-wave backgrounds which can be obtained as a Penrose limit on the $\ads_5\times S^5$ and $\ads_4\times \C P^3$ backgrounds has been performed in~\cite{Grignani:2009ny}.}, which is our starting point, has been worked out in \cite{Grignani:2008is} and discussed extensively in~\cite{Astolfi:2008ji,Astolfi:2009qh}, thus here we will report it only in the appendix \ref{app:penrose}.

%

\subsection{The procedure}

The complete Lagrangian for the type IIA Green-Schwarz (GS) string in $\ads_4\times \C P^3$ has been derived in~\cite{Gomis:2008jt, Grassi:2009yj} using a superspace construction. For convenience we report the plane-wave Lagrangian in appendix \ref{app:plane_wave}, while the complete interacting Lagrangian up to four-field terms can be found in \cite{Astolfi:2009qh}. Let us here only briefly summarize the procedure we used to derive the near-plane-wave corrections.
\begin{itemize}
\item The starting point is a plane-wave string configuration obtained taking a Penrose limit of $\ads_4\times\C P^3$ as illustrated in appendix \ref{app:penrose}.
\item The light-cone gauge must be fixed in order to remove the unphysical bosonic degrees of freedom, namely
\be
\label{lc_gf}
t(\tau\, ,\sigma) &=& c \tau\,,
\qquad
{\partial \CL \over \partial \dot v}= \text{constant}\,, \qquad {\partial \CL \over \partial v'}=0
\ee
where $v$ plays the role of the light-cone coordinate $x^-$ and%
\footnote{The constant c is computed by imposing $\int_0^{2\pi} {d\sigma\over 2\pi} p_{v}=\frac{2 J}{R^2}$.}
 $c={4J\over R^2}= {J\over \pi\sqrt{2\lambda}}\equiv {1\over\pi \sqrt{2\lambda'}}$.

\item The fermions for the type IIA superstring are real Majorana-Weyl spinors with 32 components: $\theta=\theta^1+\theta^2$ with $\Gamma_{11}\theta^1=\theta^1$ and $\Gamma_{11}\theta^2=-\theta^2$. However, since the $\ads_4\times\C P^3$ background preserves only 24 supercharges out of the initial 32~\cite{Nilsson:1984bj}, in order to work with the fermionic d. o. f. corresponding to the unbroken supersymmetries, namely the 24 physical fermionic d.o.f., the appropriate $\kappa$-symmetry gauge must be fixed. This was extensively analyzed in Ref.~\cite{Astolfi:2009qh} and here we adopt the same gauge choice, cf. \eqref{k_symm_gauge} and subsequent discussion.
\item The world-sheet metric can be fixed to a Minkowski metric only to leading order and ${1\over R}$-corrections to the world-sheet metric are allowed, since in general the world-sheet conformal gauge does not commute with the equations of motion for $v$. The world-sheet metric should then be derived as a series expansion in powers of $1/R$.
\item The Virasoro constraints can be used to solve for $\dot v$ and $v'$ order by order in ${1\over R}$. These should also be used to compute the corrections to the world-sheet metric.
\item The gauge fixed Lagrangian, $\CL_{\rm gf}= \CL-{\partial \CL\over \partial \dot v }\dot v$,  is obtained using the solutions for $\dot v$ and $v'$ and has the following expansion in powers of $1/R$
\begin{equation}\label{lgf}
\CL_{\rm gf} = \CL_{2,B} + \CL_{2,F} + \frac{1}{R} ( \CL_{3,B} +
\CL_{3,BF} ) + \frac{1}{R^2} ( \CL_{4,B} + \CL_{4,BF} + \CL_{4,F} )
+ \CO(R^{-3})
\end{equation}
where we have separated purely bosonic $B$, purely fermionic $F$ and mixed terms $BF$.
\end{itemize}
After the light-cone gauge fixing \eqref{lc_gf}, we are left with the 8 transverse coordinates
\be
\label{8_bosonic_coord}
u_1\,, u_2\,, u_3\,, u_4\,, x_1\,, y_1\,, x_2\,, y_2
\ee
where the first three are the transverse directions of $\ads_4$, while the last five are the transverse directions of $\C P^3$, cf. appendix \ref{app:penrose}.

>From the expression \eqref{lgf} follows an analogous expansion for the light cone Hamiltonian
\begin{equation}
\label{hgf}
H_{\rm gf} = H_{2,B} + H_{2,F} + \frac{1}{R} ( H_{3,B} +
H_{3,BF} ) + \frac{1}{R^2} ( H_{4,B} + H_{4,BF} + H_{4,F} )
+ \CO(R^{-3})
\end{equation}
%


\subsection{Mode expansions and plane-wave spectrum}

The solutions to the classical e.o.m. for the bosonic fields can be written as, cf. appendix \ref{app:plane_wave},
\begin{equation}
\label{zmode} z_a(\tau,\sigma) = 2 \sqrt{2} \, e^{i\frac{ c
\tau}{2}} \sum_{n} \frac{1}{\sqrt{\omega_n}} \Big[ a_n^a
e^{-i ( \omega_n \tau - n \sigma ) } -  (\tilde{a}^a)^\dagger_n e^{i
( \omega_n \tau - n \sigma ) } \Big]
\end{equation}
\begin{equation}
u_i (\tau,\sigma ) = i \frac{1}{\sqrt{2}} \sum_{n}
\frac{1}{\sqrt{\Omega_n}} \Big[ \hat{a}^i_n e^{-i ( \Omega_n \tau -
n \sigma ) } - (\hat{a}^i_n)^\dagger e^{i ( \Omega_n \tau - n \sigma
) } \Big]
\end{equation}
where we defined $z_a(\tau,\sigma)=x_a(\tau,\sigma)+iy_a(\tau,\sigma)$. The eight physical bosonic degrees of freedom are split into two families: the ``light'' bosons, whose dispersion relation is given by $\omega_n=\sqrt{\frac{c^2}{4}+n^2}$ and which corresponds to the fields $(x_1\,, y_1\,, x_2\,,y_2)$  and the  ``heavy'' bosons, corresponding to $(u_1\,, u_2\,, u_3\,, u_4)$, whose dispersion relation is given by $\Omega_n=\sqrt{c^2+n^2}$.\footnote{These modes are ``heavy" since their world-sheet mass is twice that of the ``light" modes.}
The canonical commutation relations are
$
[x_a(\tau,\sigma),\Pi_{x_b}(\tau,\sigma')] = i\delta_{ab} \delta
(\sigma-\sigma')$, $[y_a(\tau,\sigma),\Pi_{y_b}(\tau,\sigma')] =
i\delta_{ab}\delta (\sigma-\sigma')$ and
$[u_i(\tau,\sigma),\Pi_{u_j}(\tau,\sigma')] = i\delta_{ij} \delta
(\sigma-\sigma')$ and they follow from
\begin{equation}
\label{comrel}
[a_m^a,(a_n^b)^\dagger] = \delta_{mn} \delta_{ab}\spa
[\tilde{a}_m^a,(\tilde{a}_n^b)^\dagger] = \delta_{mn}
\delta_{ab}\spa [\hat{a}^i_m,(\hat{a}^j_n)^\dagger] = \delta_{mn}
\delta_{ij}
\end{equation}
%


For the fermions, the mode expansions which follow from the e.o.m. derived from the plane-wave Lagrangian are
\begin{equation} \label{modepsiplus}
\psi_{+,\alpha} =  \frac{\sqrt{2\alpha'}}{\sqrt{c}} \sum_{n} \Big[ f^+_n d_{n,\alpha}
e^{-i ( \omega_n \tau - n \sigma ) } - f^-_n d^\dagger_{n,\alpha}
e^{i ( \omega_n \tau - n \sigma ) } \Big]
\end{equation}
\begin{equation} \label{modepsiminus}
\psi_{-,\alpha} =\frac{\sqrt{\alpha'}}{\sqrt{c}}  \big( e^{- \frac{c}{2} \Gamma_{56}
\tau} \big)_{\alpha \beta} \sum_{n} \Big[ - g^-_n b_{n,\beta}
e^{-i ( \Omega_n \tau - n \sigma ) } + g^+_n b^\dagger_{n,\beta}
e^{i ( \Omega_n \tau - n \sigma ) } \Big]
\end{equation}
where the fermionic fields $\psi_{\pm}$ are defined in \eqref{psipm} and the functions $f^\pm_n$ and $g^\pm_n$ are given by
\begin{equation}\label{effeg}
f^\pm_n = \frac{\sqrt{\omega_n+n} \pm
\sqrt{\omega_n-n}}{2\sqrt{\omega_n}} \spa g^\pm_n =
\frac{\sqrt{\Omega_n+n} \pm \sqrt{\Omega_n-n}}{2\sqrt{\Omega_n}}
\end{equation}
We see that also the fermions split into four ``light" (the $d$'s) and four ``heavy" (the $b$'s) d.o.f. subject to the conditions
\begin{equation}
\CP_+ d_n = d_n \spa \Gamma_{11} d_n = d_n \spa \CP_- b_n = b_n \spa \Gamma_{11} b_n = b_n
\end{equation}
and obey the anti-commutation relations
\begin{equation}
\{ d_{m,\alpha} , d^\dagger_{n,\beta} \} = \delta_{mn }
\left(\frac{1+\Gamma_{11}}{2} \CP_+\right)_{\alpha\beta} \spa \{ b_{m,\alpha} , b^\dagger_{n,\beta} \} =
\delta_{mn } \left(\frac{1+\Gamma_{11}}{2} \CP_-\right)_{\alpha\beta}
\end{equation}
For our conventions on $\Gamma$ matrices see Appendix~\ref{app:gamma}.
>From this one obtains $\{ \psi_{\alpha} (\tau,\sigma) , \psi_{\beta}^* (\tau,\sigma') \} = \frac{2\pi \alpha'}{c} (\frac{1+\Gamma_{11}}{2}( \CP_- + 2 \CP_+) )_{\alpha\beta} \delta ( \sigma-\sigma' )$, which in terms of the fermionic momenta \eqref{def_rho} can be written as
\begin{equation}
\label{anticommut}
\{ \psi_{\alpha} (\tau,\sigma) , \rho_{\beta} (\tau,\sigma') \} = - 2\pi i \alpha' \left(\frac{1+\Gamma_{11}}{2} ( \CP_- + \CP_+) \right)_{\alpha\beta} \delta(\sigma-\sigma')
\end{equation}
The GS string $\sigma$-model we are considering has first and second class constraints which require a Dirac procedure in order to be properly treated. However one of the main achievements of \cite{Astolfi:2009qh} was to perform a fermionic field redefinition which allows one to use the standard canonical anticommutation relations \eqref{anticommut} up to the order considered here.  \\

Plugging the mode expansion of the fermionic and bosonic fields in \eqref{CH2B} and \eqref{CH2F} one obtains the mode expansion for the pp-wave bosonic and fermionic Hamiltonians
\begin{equation}
c H_{2,B} =  \sum_{n} \left[\sum_{i=1}^4\Omega_n
\hat{N}^i_n+\sum_{a=1}^2 \left(\omega_n -
\frac{c}{2}\right) M_n^a +\sum_{a=1}^2 \left(\omega_n+
\frac{c}{2}\right) N_n^a \right]\label{penspectrum}
\end{equation}
\begin{equation}
\label{fermppwave}
c H_{2,F} = \sum_{n} \left[ \sum_{f=1}^4 \omega_n F_n^{(f)} +
\sum_{f=5}^6 \left( \Omega_n + \frac{c}{2} \right) F_n^{(f)} +
\sum_{f=7}^8 \left( \Omega_n - \frac{c}{2} \right) F_n^{(f)} \right]
\end{equation}
The bosonic number operators are $\hat{N}^i_n = (\hat{a}^i_n)^\dagger
\hat{a}^i_n$, with $i=1,\dots,4$, $M_n^a = (a^a)^\dagger_n a^a_n$ and $N_n^a =
(\tilde{a}^a)^\dagger_n \tilde{a}_n^a$ with $a=1,2$, while for the fermions we have $F^{(f)}_n= d_{n,\alpha}^\dagger d_{n,\alpha}$ for $f=1,\ldots ,4$, and
$F_n^{(f)} = b_{n,\alpha} ^\dagger b_{n,\alpha}$ for $f=5,\ldots,8$.
Finally, the level-matching condition is
\begin{equation}
\label{levelmbf} \sum_{n}n \left[\sum_{i=1}^4
\hat{N}^i_n+\sum_{a=1}^2 \left(M_n^a + N_n^a\right) +\sum_{f=1}^8
F^{(f)}_n\right]
 = 0
\end{equation}


\section{The near plane-wave spectrum in the $\grSU(2)\times \grSU(2)$ sector}
\label{sec:string_finitesize_en}

In this section we compute corrections to the energy of certain states in the $\grSU(2)\times \grSU(2)$ sector of type IIA string theory on $\ads_4\times \C P^3$ in a near pp-wave expansion. To illustrate the computation we consider a two oscillator state of the form
\begin{equation}\label{s}
|s\rangle = (a_n^1)^\dagger (a_{-n}^1)^\dagger|0\rangle
\end{equation}
This is a state in which both oscillators are in the same $\grSU(2)$ sector. Then we will generalize the result to the case in which the two oscillators are in different $\grSU(2)$'s and to the case in which we have a state with a generic number of oscillators. Moreover, in Sec.~\ref{sec:outside_SU2xSU2} we will consider the correction to the energy of states outside the $\grSU(2)\times \grSU(2)$ sector.

Note that the state~\eqref{s} has a degenerate spectrum, at the pp-wave level, with the state
\begin{equation}\label{t}
|t\rangle =\frac{1}{2}\left( (a^1_n)^\dagger (a^2_{-n})^\dagger+(a^1_{-n})^\dagger (a^2_{n})^\dagger\right)|0\rangle
\end{equation}
where there is an oscillator in each of the two $\grSU(2)$'s. However, the two states are not mixed when perturbations to the pp-wave Hamiltonian are included~\cite{Astolfi:2008ji}, therefore, one can use perturbation theory for non-degenerate states.

At first order in perturbation theory the cubic Hamiltonian given in Eqs.~\eqref{CH3B} and \eqref{finalH3BF} does not contribute to the corrections to the energy of the state
$|s\rangle$ since its mean value on this state vanishes. So the first non-trivial correction to the energy sets in at order
$\frac{1}{R^2}$.
At this order, there are two contributions, one coming from the
second perturbative order generated by the cubic Hamiltonian~\eqref{CH3B} and \eqref{finalH3BF}
and one that arises from the first perturbative order by taking
the mean value of the quartic Hamiltonian \eqref{CH4B},  \eqref{pH4F}  and \eqref{finalH4BF} on the state
$|s\rangle$.
We can thus write
\begin{equation}\label{energycorrection}
E_{s}^{(2)}=\frac{1}{R^2}\left(\sum_{|i\rangle \neq |s\rangle}\frac{\left|\langle i|H_{3}|s\rangle\right|^2}{E^{(0)}_{|s\rangle}-E^{(0)}_{|i\rangle}}+\langle s |H_{4}|s\rangle \right)
\end{equation}
where $|i\rangle$ is an intermediate state with zeroth order energy
$E^{(0)}_{|i\rangle}$.

\subsection{The term $\sum_{|i\rangle \neq |s\rangle}\frac{\left|\langle i|H_{3}|s\rangle\right|^2}{E^{(0)}_{|s\rangle}-E^{(0)}_{|i\rangle}}$}
\label{sec:cubic_H}

Let us consider the first term in Eq.~\eqref{energycorrection}.
It is easy to see that the total contribution can be divided into two separate contributions, namely we can write
\begin{equation}
\sum_{|i\rangle \neq |s\rangle}\frac{\left|\langle i|H_{3}|s\rangle\right|^2}{E^{(0)}_{|s\rangle}-E^{(0)}_{|i\rangle}}=\sum_{|i\rangle \neq |s\rangle}\frac{\left|\langle i|H_{3,B}|s\rangle\right|^2}{E^{(0)}_{|s\rangle}-E^{(0)}_{|i\rangle}}+\sum_{|i\rangle \neq |s\rangle}\frac{\left|\langle i|H_{3,BF}|s\rangle\right|^2}{E^{(0)}_{|s\rangle}-E^{(0)}_{|i\rangle}}
\label{fullh3tot}
\end{equation}
where $H_{3,B}$ and $H_{3,BF}$ are given in \eqref{CH3B} and \eqref{finalH3BF} respectively.

The relevant part of the cubic Hamiltonian contributing to the first term in
 \eqref{fullh3tot}, written in terms of oscillators, is given by~\cite{Astolfi:2008ji, Astolfi:2009qh}
\begin{eqnarray}\label{hc1osc}
&& H_{3,B} =\frac{i}{c\sqrt{2}}\sum_{m,\,l,\,r}\bigg\{
\frac{\delta(m+l+r)(\hat a_{-r}^4)^\dagger}{\sqrt{\omega_m\omega_l\Omega_r}}
\left[\left(\omega_m-\frac{c}{2}\right)\left(\omega_l-\frac{c}{2}\right)+ml\right]\cdot\cr&&
\left[(a_{-m}^2)^\dagger
(a_l^2)-(a_{-m}^1)^\dagger (a_l^1)\right]+\left[\left(\omega_m+\frac{c}{2}\right)\left(\omega_l-\frac{c}{2}\right)-ml\right]\cdot\cr&&\left[\left(\tilde a_m^1\right)\left(a_l^1\right)+\left(\tilde a^1_{-m}\right)^\dagger \left(a^1_{-l}\right)^\dagger-\left(\tilde a_m^2\right)\left(a_l^2\right)+\left(\tilde a^2_{-m}\right)^\dagger \left(a^2_{-l}\right)^\dagger\right]\bigg\}
\end{eqnarray}
This Hamiltonian produces divergent results at second order in perturbation theory~\cite{ Astolfi:2008ji} therefore it is natural to handle the sums over mode numbers by introducing cutoffs, and removing the cutoffs only at the end of the calculations. After all, this is the way infinite sums are defined.
It is then important to notice, that the form of the cubic interacting Hamiltonian~\eqref{hc1osc} uniquely fixes the possible choices of the cutoffs on the sums on mode numbers. Under the quite natural assumption that all the light modes have the same cutoff $N$ for the level numbers, and that all the heavy modes have the same cutoff $M$ for the level numbers
 the form of the interaction naturally suggests $M = 2N$. As can be seen from \eqref{penspectrum}, $\hat a^4$ is the oscillator of a heavy mode whereas $a$ and $\tilde a$ are oscillators of light modes.
Consequently, the Hamiltonian \eqref{hc1osc} contains one heavy oscillator and two light oscillators, therefore if the sums over $m$ and $l$ have cutoff $N$, the sum over $r$ has cutoff $2N$.

 The same argument holds also in the case of the mixed bosonic and fermionic cubic Hamiltonian entering in the computation of the second term in~\eqref{fullh3tot}.
Thus we see that the string theory interacting Hamiltonian naturally selects the regularization prescription in agreement with the prediction coming from the algebraic curve approach \cite{Gromov:2008bz,Leoni:2010tb}. In \cite{Bandres:2009kw} a similar prescription was proposed in order to obtain a result consistent with the algebraic curve approach. The motivation behind this prescription is essentially that heavy excitations are not fundamental but rather bound states of two light fundamental modes~\cite{Zarembo:2009au}. Here we see that this arises from the form of the interacting Hamiltonian, leading to a prescription for summing over the world-sheet frequencies which distinguishes between heavy and light modes.

The intermediate states that one has to consider in the computation of the first term in
 \eqref{fullh3tot} are of the form~\footnote{We thank K. Zarembo for discussions on this point.}
$| i_1 \rangle =(a^4_{-p-q})^\dagger
(a^1_{p})^\dagger(a^1_{q})^\dagger|0\rangle$ and $| i_2 \rangle =(a^4_{-p-q-r-s})^\dagger
(a^1_{p})^\dagger(a^1_{q})^\dagger(a^1_{r})^\dagger(\tilde a^1_{s})^\dagger| 0 \rangle $.
We find
\begin{eqnarray}
\label{samesu2spectrum2part1}
\sum_{|i_1\rangle\neq |s\rangle}\frac{\left|\langle i_1|H_{3,B}|s\rangle\right|^2}{E^{(0)}_{|s\rangle}-E^{(0)}_{|i_1\rangle}}=\CS_1-
\frac{\left[\left(\omega_{n}-\frac{c}{2}\right)^2+n^2\right]^2}{c\,\omega^2_{n}\Omega^2_{2n}}
-\frac{\left[\left(\omega_{n}-\frac{c}{2}\right)^2-n^2\right]^2}{c^3\omega^2_{n}}
\end{eqnarray}
\begin{eqnarray}
\label{samesu2spectrum2part2}
\sum_{|i_2\rangle\neq |s\rangle}\frac{\left|\langle i_2|H_{3,B}|s\rangle\right|^2}{E^{(0)}_{|s\rangle}-E^{(0)}_{|i_2\rangle}}&&=\CS_2
\end{eqnarray}
where we divided by the appropriate numeric factors in order to avoid overcounting and where we have separated the finite contributions from the following logarithmically divergent sums
\begin{equation}
\CS_1=\frac{1}{2c}\sum_{q=-N}^{N}\left[\frac{\left[\left(\omega_{q}-\frac{c}{2}\right)\left(\omega_{n}-\frac{c}{2}\right)+q n\right]^2}
{\omega_{q}\omega_n\Omega_{q+n}\left(\omega_{n}-\omega_q-\Omega_{q+n}\right)}+\left(n\leftrightarrow -n\right)\right]
\label{s1}
\end{equation}
\begin{equation}
\CS_2=-\frac{1}{2c}\sum_{q=-N}^{N}\left[\frac{\left[\left(\omega_{q}+\frac{c}{2}\right)\left(\omega_{n}-\frac{c}{2}\right)-q n\right]^2}
{\omega_{q}\omega_n\Omega_{q+n}\left(\omega_{q}+\omega_n+\Omega_{q+n}\right)}+\left(n\leftrightarrow -n\right)\right]
\label{s2}
\end{equation}
After the contractions among the oscillators in the Hamiltonian and in the intermediate and external states are performed, we are left only with a cutoff $N$ since, at the end of the calculation, we are left with only the sum over a light mode.
As we will see, similar divergent sums are also generated by the second term in Eq.~\eqref{fullh3tot}. However, these divergences can not cancel among each other since second order corrections in perturbation theory are always negative. We will show that including the appropriate normal ordering prescription in the computation of $\langle s|H_{4}|s\rangle$ will produce a finite result for the energy.

Combining \eqref{samesu2spectrum2part1} and \eqref{samesu2spectrum2part2} we obtain
\begin{eqnarray} \label{fullH3BB}
&&\sum_{|i_1\rangle\neq |s\rangle}\frac{\left|\langle i_1|H_{3,B}|s\rangle\right|^2}{E^{(0)}_{|s\rangle}-E^{(0)}_{|i_1\rangle}}+\sum_{|i_2\rangle\neq |s\rangle}\frac{\left|\langle i_2|H_{3,B}|s\rangle\right|^2}{E^{(0)}_{|s\rangle}-E^{(0)}_{|i_2\rangle}}=\cr&&=\CS_1+\CS_2-
\frac{\left[\left(\omega_{n}-\frac{c}{2}\right)^2+n^2\right]^2}{c\,\omega^2_{n}\Omega^2_{2n}}
-\frac{\left[\left(\omega_{n}-\frac{c}{2}\right)^2-n^2\right]^2}{c^3\omega^2_{n}}
\end{eqnarray}

Let us now consider the second term in Eq.~\eqref{fullh3tot}.
In terms of oscillators $H_{3,BF}$ can be written as follows
\begin{eqnarray}
\label{H3BFoscillators}
 H_{3,BF} &= &
 \frac{1}{8 \sqrt{2} R c}\Bigg\{\sum_{m, p , s} \frac{\delta(m+p+s)m \,a^\dagger_{-m}}{\sqrt{\omega_m}}\bigg[
 2\left(g_p^- f_s^-+g_p^+ f_s^+\right)
\left(\left(i \Gamma_9-2 \Gamma_4\right)\left(\Gamma_5+i \Gamma_6\right)\right)_{\gamma, \delta}
\cr
&+& \left(g_p^- f_s^- - g_p^+ f_s^+\right)
\left(\left(-i \Gamma_0+3 i\Gamma_{12340}+2 \Gamma_4\right)\left(\Gamma_5+i \Gamma_6\right)\right)_{\gamma, \delta}
\bigg] \cr
 &+& \sum_{m, p , s} \frac{\delta(m+p+s)a^\dagger_{-m}\left(\omega_m-\frac{c}{2}\right)}{\sqrt{\omega_m}}\bigg[
  2\left(g_p^- f_s^++g_p^+ f_s^-\right)\left(\left(i \Gamma_9-2 \Gamma_4\right)\left(\Gamma_5+i \Gamma_6\right)\right)_{\gamma, \delta}
 \cr
&+& \left(g_p^- f_s^+-g_p^+ f_s^-\right)\left(\left(-i \Gamma_9+3 i \Gamma_{12349}+2 \Gamma_4\right)\left(\Gamma_5+i \Gamma_6\right)\right)_{\gamma, \delta}
\bigg]\Bigg\} b_{p, \gamma} d_{s, \delta}
+ (\rm{c.c.}) +\dots\cr &&
\end{eqnarray}
where $f^{\pm}_n$ and $g^{\pm}_n$ are defined in~\eqref{effeg} and the dots stand for terms that are irrelevant in the calculation of the energy of the state $|s\rangle$. The sums over the modes $m$ and $s$ corresponding to oscillators $a$ and $d$ have cutoff $N$, these are in fact light modes as can be seen from \eqref{penspectrum} and \eqref{fermppwave}. Consequently, the sum over the heavy mode $p$, corresponding to the oscillator $b$  in  \eqref{penspectrum} and \eqref{fermppwave}, has cutoff $2N$, it is a heavy mode.

The intermediate states that we should consider in this case are $| i_1\rangle = (a^1_r)^\dagger d^\dagger_{q,\alpha} b^\dagger_{-r-q, \beta}| 0\rangle$ and $| i_2\rangle = (a^1_r)^\dagger (a^1_t)^\dagger (a^1_u)^\dagger d^\dagger_{q,\alpha} b^\dagger_{-u-t-r-q, \beta}| 0\rangle$.
We find
\be
\label{sum_h3bf_squared1bis}
\sum_{| i_1 \rangle \ne | s \rangle}\frac{\big|\langle s |H_{3,BF}| i_1\rangle\big|^2 }{E^{(0)}_{| s \rangle}-E^{(0)}_{| i_1 \rangle}}+\sum_{| i_2 \rangle \ne | s \rangle}\frac{\big|\langle s |H_{3,BF}| i_2\rangle\big|^2 }{E^{(0)}_{| s \rangle}-E^{(0)}_{| i_1 \rangle}} =
\CS_3+\CS_4
\ee
where
\begin{eqnarray}
\CS_3=&&\sum_{q=-N}^N\left[\frac{1}{32 \,c \, \omega_n\omega_q\Omega_{q+n}\left(\omega_n-\omega_q-\Omega_{q+n}\right)}\left[18\left(\omega_n-\frac{c}{2}\right)^2\left[2 \omega_q\Omega_{q+n}+2 c \omega_q\right.\right.\right.\nn\\ &&\left.\left.\left.
- c\Omega_{q+n}- c^2\right]+2 n^2\left[10 \omega_q \Omega_{q+n}-8 q\left(n+q\right)+5 c^2-6 c \omega_q-3 c\Omega_{n+q}\right]\right.\right.\nn\\ &&\left.\left.
+24 n\left(\omega_{n}-\frac{c}{2}\right)\left[-2\left(n+q\right)\omega_q+q \Omega_{q+n}+2 c q+ c n\right]\right]+ \left(n\leftrightarrow -n\right)\right]
\label{s3}
\end{eqnarray}
and
\begin{eqnarray}
\CS_4=&&-\sum_{q=-N}^N\bigg[\frac{1}{32 \,c \,\omega_n\omega_q\Omega_{q+n}\left(\omega_n+\omega_q+\Omega_{q+n}\right)}\bigg\{18\left(\omega_n-\frac{c}{2}\right)^2\big[2 \omega_q\Omega_{q+n}\cr&&-2 c \omega_q
+ c\, \Omega_{q+n}- c^2\big]
+2 n^2\left[10 \omega_q \Omega_{q+n}-8 q\left(n+q\right)+5 c^2+6 c \omega_q+3 c\Omega_{n+q}\right]
\nn \\ &&
+24 n\left(\omega_{n}-\frac{c}{2}\right)\left[2\left(n+q\right)\omega_q-q \Omega_{q+n}+2 c q+ c n\right]\bigg\}+ \left(n\leftrightarrow -n\right)\bigg]\cr&&
\label{s4}
\end{eqnarray}

Adding together the results~\eqref{fullH3BB} and \eqref{sum_h3bf_squared1bis} we find
\begin{eqnarray}
\label{cubicspectrum}
\sum_{|i\rangle\ne | s \rangle}\frac{\left|\langle i|H_{3}|s\rangle\right|^2}{E^{(0)}_{|s\rangle}-E^{(0)}_{|i\rangle}}&=&
\CS_1+\CS_2+\CS_3+\CS_4-
\frac{\left[\left(\omega_{n}-\frac{c}{2}\right)^2+n^2\right]^2}{4\,c\,\omega^4_{n}}
-\frac{\left[\left(\omega_{n}-\frac{c}{2}\right)^2-n^2\right]^2}{c^3\omega^2_{n}}\cr &&
\end{eqnarray}
The sums $\CS_i$ are logarithmically divergent. We will show that these divergences are canceled when including the contribution of the contact term in Eq.~\eqref{energycorrection} thus giving a finite result for the energy.

\subsection{The term $\langle s |H_{4}|s\rangle$}
\label{sec:H4}

Now we want to compute $\langle s |H_{4}|s\rangle$, where $H_4=H_{4,B}+H_{4,F}+H_{4,BF}$. The various expressions in terms of fields for these Hamiltonians were computed in~\cite{Astolfi:2009qh} and we have reproduced them in App.~\ref{app:Fermionsandmore}, Eqs. \eqref{CH4B}, \eqref{pH4F} and \eqref{finalH4BF} respectively. As we saw in the previous section, for the first term in Eq.~\eqref{energycorrection} we get a divergent result, therefore to obtain a finite result for the energy, this divergence must be canceled by the second term~\footnote{A similar cancelation of divergences between a cubic term and a quartic ``contact" term was already shown to happen in~\cite{Grignani:2005yv,Grignani:2006en,Astolfi:2008yw}.} . In order for this cancelation to happen the quartic Hamiltonian cannot be normal ordered as it is for type IIB superstring on $\ads_5 \times \sphere^5$~\cite{Callan:2003xr}, otherwise its mean value would just be finite. We shall thus introduce for it an appr!
 opriate and consistent normal ordering prescription. Choosing the most natural normal ordering prescription, we will show that all the divergences cancel leaving a finite result for the energy.
We will discuss in Section~\ref{sec:no} more generally the question of how to properly normal order a quantity when promoting oscillators to quantum operators. We will see that requiring that the pp-wave algebra is not affected by normal ordering constants and that the spectrum of string states is finite, will fix uniquely the normal ordering prescription.

In deriving the  normal ordering prescription, we will also assume that the vacuum is a protected state. This automatically implies that the term $H_{4,F}$ does not contribute to the energy of the state $|s\rangle$, since if it would, it would also change the vacuum energy. Thus we can write
\be
\label{h4tot}
\langle s |H_{4}|s\rangle=\langle s | H_{4,B}| s\rangle + \langle s | H_{4,BF}| s\rangle
\ee
and we will consider the two terms separately. Here, for simplicity, we will not report the explicit expressions for $H_{4,B}$ and $H_{4,BF}$ in terms of oscillators. They can be derived from the Eqs.~\eqref{CH4B} and \eqref{finalH4BF} in App.~\ref{app:Fermionsandmore}, but are too complicated to display here.

Using the totally symmetric prescription, namely  choosing that the normal ordering constants for the two oscillator terms are equal to $1/2$ and those for the normal ordering of terms with 4 oscillators of the same kind are equal to $1/6$ (see Section~\ref{sec:no} for details), for the first term in \eqref{h4tot} we get
\begin{equation}
\label{H4BBsym}
  \langle s | H_{4,B} | s\rangle_{\rm sym} =-\frac{2\left[\left(\omega_n-c\right)\left(4n^2-c^2\right)-c^2
\omega_n\right]}{c^3 \omega_n}+
\CS_5+\CS_6
\end{equation}
where the first term is the contribution of the normal ordered terms, which was already computed in~\cite{Astolfi:2008ji}, and the rest of the terms are obtained employing the normal ordering prescription described in Section~\ref{sec:no}. We get that  $\CS_5$ and $\CS_6$ are the regularized quadratically divergent sums
\begin{equation}
\label{s5}
\CS_5= -\sum_{q=-N}^N\frac{1}{c^3\,\omega_n\,\omega_q}
\left[8 q^2 \left(n^2+(\omega_n-\frac{c}{2})^2\right)
 +c^2\left(n^2+3(\omega_n-\frac{c}{2})^2+4 c(\omega_n-\frac{c}{2})\right)\right]
\end{equation}
and
\begin{equation}
\label{s6}
\CS_6=-\sum_{q=-2N}^{2N}\frac{1}{c^3\,\omega_n\,\Omega_q}
 \left[8 q^2 \left[n^2+\left(\omega_n-\frac{c}{2}\right)^2\right]+2 n^2c^2\right]
\end{equation}
The cutoffs have been chosen according to the natural requirement, imposed by the form of the cubic Hamiltonian, that the cutoff on heavy modes is twice that on light modes.

Using the symmetric prescription for the normal ordering, the computation of the second term in Eq.~\eqref{h4tot}
gives
\begin{eqnarray}
\label{H4BFsym}
\langle s | H_{4,BF} | s\rangle_{\rm sym}
= \CS_7+\CS_8
\end{eqnarray}
where
\begin{equation}
\label{s7}
\CS_7= \sum_{q=-N}^N\frac{1}{4\,c^3\, \omega_n\,\omega_q}\bigg[ 32 q^2 \left( n^2+\left(\omega_n-\frac{c}{2}\right)^2\right)
+c^2\left(11 n^2-9\left(\omega_n-\frac{c}{2}\right)^2\right)\bigg]
\end{equation}
and
\begin{equation}
\label{s8}
\CS_8=\sum_{q=-2N}^{2N}\frac{1}{2\,c^3\, \omega_n\,\Omega_q}\bigg[ 16 q^2 \left( n^2+\left(\omega_n-\frac{c}{2}\right)^2\right)
+c^2\left(11 n^2+9\left(\omega_n-\frac{c}{2}\right)^2\right)\bigg]
\end{equation}

Adding together the results~\eqref{H4BBsym} and \eqref{H4BFsym} we obtain
\begin{eqnarray} \label{quarticspectrum}
\langle s |H_{4}|s\rangle_{\rm sym} &=&\CS_5+\CS_6+\CS_7+\CS_8 -\frac{2\left[\left(\omega_n-c\right)\left(4n^2-c^2\right)-c^2
\omega_n\right]}{c^3 \omega_n}
\end{eqnarray}
The result \eqref{quarticspectrum} is only logarithmically divergent, even if each sum $\CS_i$ $i=5,\dots,8$ is quadratically divergent. Quadratic and linear divergences cancel between the contributions coming from  $H_{4,B}$ and $H_{4,BF}$, only the logarithmic divergence remains,
and has the same form but the opposite sign of the one coming from the  $\sum_{|i\rangle \neq |s\rangle}\frac{\left|\langle i|H_{3}|s\rangle\right|^2}{E^{(0)}_{|s\rangle}-E^{(0)}_{|i\rangle}}$ term.


\subsection{String energy spectrum at order $1/R^2$}
\label{spectrum_reload}

We can now compute the $1/R^2$ correction to the energy of the state $|s\rangle$ putting together the results~\eqref{cubicspectrum} and \eqref{quarticspectrum}. We obtain
\begin{equation} \label{senergy}
E_s^{(2)}=-\frac{8\,n^2\left[\left(\omega_n-\frac{c}{2}\right)^2-\frac{c^2}{2}\right]}{R^2c^3\omega_n^2}
+\CS(n)
 \end{equation}
where we introduce the sum $\CS(n)$ defined as
\begin{equation}
\label{esse}
\CS(n)\equiv\frac{\sum_{i=1}^8\CS_i}{R^2}=\frac{1}{2\,c R^2\,\omega_n}
\left[\sum_{q=-N}^{N}\frac{9c\,\omega_n-\frac{9}{2}c^2-8n^2}{\Omega_{q+n}}-\sum_{q=-2N}^{2N}\frac{9c\,\omega_n
-\frac{9}{2}c^2-16n^2}{\Omega_{q}}-\sum_{q=-N}^{N}\frac{8 n^2}{\omega_q}\right]
\end{equation}
As anticipated, now we can see that the result is free of divergences. In the next section, we will discuss different possible ways of regularizing the sum $\CS$, then use the one imposed by the cubic interaction Hamiltonian and
derive the finite result arising from it.

Doing the computation for the state $|t\rangle$ in an analogous way, we find
\begin{equation} \label{tenergy}
E_t^{(2)}=-\frac{8\,n^2\left(\omega_n-\frac{c}{2}\right)^2}{R^2c^3\omega_n^2}+\CS(n)
\end{equation}
Again, the result is free of divergences.

\paragraph{States with arbitrary number of $SU(2)\times SU(2)$ oscillators.}

We can repeat the computation also for a generic state with $K$ oscillators in one $SU(2)$ and $K'$ in the other $SU(2)$
\begin{equation} \label{state}
| K,K'\rangle\equiv\left(a^1_{n_1}\right)^\dagger\dots \left(a^1_{n_K}\right)^\dagger \left(a^2_{n_{1'}}\right)^\dagger\dots \left(a^2_{n_{K'}}\right)^\dagger | 0\rangle~~,~~
\end{equation}
obeying the level matching condition
$
\sum_{i=1}^K n_i+\sum_{i'=1}^{K'} n_{i'}=0$, where all the $n_i$  and $n_{i'}$ are different.
We find
\begin{eqnarray} \label{energygeneralc}
&&E^{(2)}_{| K,K'\rangle}=  \frac{1}{2R^2 c} \sum_{i=1}^K \sum_{j\ne i} \frac{\left[\left(\omega_{n_i}-\frac{c}{2}\right)\left(\omega_{n_j}-\frac{c}{2}\right)- n_i n_j\right]^2}{\omega_{n_i}\omega_{n_j}\Omega_{n_i-n_j}\left(\omega_{n_i}-\omega_{n_j}-\Omega_{n_i-n_j}\right)} +\left(i\leftrightarrow i'~,~ j\leftrightarrow j'~,~ K\leftrightarrow K'\right) \cr &
-&\frac{1}{2R^2 c^3 }\sum_{i=1}^K \sum_{j\ne i} \frac{\left[\left(\omega_{n_i}-\frac{c}{2}\right)^2- n_i^2 \right] \left[\left(\omega_{n_j}-\frac{c}{2}\right)^2- n_j^2 \right]}{\omega_{n_i}\omega_{n_j}} +\left(i\leftrightarrow i'~,~ j\leftrightarrow j'~,~ K\leftrightarrow K'\right) \cr &
+&\frac{1}{R^2c^3}\sum_{i=1}^K \sum_{i'=1}^{K'} \frac{\left[\left(\omega_{n_i}-\frac{c}{2}\right)^2- n_i^2 \right] \left[\left(\omega_{n_{i'}}-\frac{c}{2}\right)^2- n_{i'}^2 \right]}{\omega_{n_i}\omega_{n_{i'}}} \cr &
+&\sum_{i=1}^K \sum_{j\ne i}\frac{1}{R^2\omega_{n_i}\omega_{n_j}}\bigg[-\frac{n_i^2+n_j^2+4n_i n_j}{c}-\frac{4 n_i^2 n_j^2}{c^3}+\frac{2 \omega_{n_i}\omega_{n_j}}{c}+\frac{2}{c^2}\left(n_i^2 \omega_{n_j}+n_j^2 \omega_{n_i}\right)\cr &+&\frac{c}{2}+\frac{4 n_i n_j\left(\omega_{n_i}-\frac{c}{2}\right)\left(\omega_{n_j}-\frac{c}{2}\right)}{c^3}-\omega_{n_i}-\omega_{n_j}\bigg]+\left(i\leftrightarrow i'~,~ j\leftrightarrow j'~,~ K\leftrightarrow K'\right)
\cr & -&\frac{1}{R^2c^3} \sum_{i=1}^K \sum_{i'=1}^{K'} \frac{1}{\omega_{n_i}\omega_{n_{i'}}}\left[\left(\omega_{n_i}-\frac{c}{2}\right)^2+n_i^2\right]\left[\left(\omega_{n_{i'}}-\frac{c}{2}\right)^2+n_{i'}^2\right]
\cr & +&\frac{4}{R^2c^3} \sum_{i=1}^K \sum_{i'=1}^{K'} \frac{1}{\omega_{n_i}\omega_{n_{i'}}} n_i n_{i'}\left(\omega_{n_i}-\frac{c}{2}\right)\left(\omega_{n_{i'}}-\frac{c}{2}\right)+\frac{1}{2}\sum^K_{i=1}\CS(n_i)+\left(i\leftrightarrow i'~,~ K\leftrightarrow K'\right)\cr &&
\end{eqnarray}
Note that the contribution coming from the sum $\CS(n)$ is diagonal in each of the $n_i$ and $n_{i'}$, thus showing that this term should be considered as arising from a finite size correction to the dispersion relation of the single magnon and not from the interaction among them. In the next Section we will give an interpretation for it.

\paragraph{States with one $\grSU(2)$ oscillator.}

In order to provide further insight whether the sum $\CS(n)$ should be ascribed as a finite size correction to the single magnon dispersion relation, we inquire what happens if one considers a single oscillator state in one $\grSU(2)$, i.e. $| s.o. \rangle = a^\dagger_n | 0 \rangle$.
In doing this we must relax the level matching condition, which would force the mode number $n$ to be vanishing.

When computing $\sum_{|i\rangle \neq |s.o.\rangle}\frac{\left|\langle i|H_{3, B}|s.o.\rangle\right|^2}{E^{(0)}_{|s.o.\rangle}-E^{(0)}_{|i\rangle}}$ the intermediate states giving a non vanishing contribution are $| i_1 \rangle =(a^4_p)^\dagger
(a^1_{q})^\dagger|0\rangle$ and $| i_2 \rangle =(a^4_{p})^\dagger
(a^1_{q})^\dagger(a^1_{r})^\dagger(\tilde a^1_{s})^\dagger| 0 \rangle$. In order to follow a consistent procedure, the level matching condition must be relaxed on the intermediate states as well as on the external state $| s.o. \rangle$.

One computes
\begin{equation} \label{s1single}
\sum_{|i_1\rangle\neq |s.o\rangle}\frac{\left|\langle i_1|H_{3,B}|s.o.\rangle\right|^2}{E^{(0)}_{|s.o.\rangle}-E^{(0)}_{|i_1\rangle}}=\hat \CS_1=\frac{1}{2c}\sum_{q=-N}^{N}\frac{\left[\left(\omega_{q}-\frac{c}{2}\right)\left(\omega_{n}-\frac{c}{2}\right)-q n\right]^2}
{\omega_{q}\omega_n\Omega_{q-n}\left(\omega_{n}-\omega_q-\Omega_{q-n}\right)}
\end{equation}
\begin{equation} \label{s2single}
\sum_{|i_1\rangle\neq |s.o.\rangle}\frac{\left|\langle i_2|H_{3,B}|s.o.\rangle\right|^2}{E^{(0)}_{|s.o.\rangle}-E^{(0)}_{|i_2\rangle}}=\hat \CS_2=\frac{1}{2c}\sum_{q=-N}^{N}\frac{\left[\left(\omega_{q}+\frac{c}{2}\right)\left(\omega_{n}-\frac{c}{2}\right)-q n\right]^2}
{\omega_{q}\omega_n\Omega_{q+n}\left(\omega_{q}+\omega_n+\Omega_{q+n}\right)}
\end{equation}

Turning then to $\sum_{|i\rangle \neq |s.o.\rangle}\frac{\left|\langle i|H_{3, BF}|s.o.\rangle\right|^2}{E^{(0)}_{|s.o.\rangle}-E^{(0)}_{|i\rangle}}$, the intermediate states that we should consider in this case are $| i_1\rangle = d^\dagger_{q,\alpha} b^\dagger_{r, \beta}| 0\rangle$ and $| i_2\rangle = (a^1_r)^\dagger (a^1_t)^\dagger d^\dagger_{q,\alpha} b^\dagger_{u, \beta}| 0\rangle$.

One computes
\begin{equation}
\sum_{|i_1\rangle\neq |s.o.\rangle}\frac{\left|\langle i_1|H_{3,BF}|s.o.\rangle\right|^2}{E^{(0)}_{|s.o.\rangle}-E^{(0)}_{|i_1\rangle}}=\hat \CS_3~,
\label{s3hat}
\end{equation}
with
\begin{eqnarray}
\hat \CS_3=&&\sum_{q=-N}^N\frac{1}{32 \,c \, \omega_n\omega_q\Omega_{q-n}\left(\omega_n-\omega_q-\Omega_{q-n}\right)}\left[18\left(\omega_n-\frac{c}{2}\right)^2\left[2 \omega_q\Omega_{q-n}+2 c \omega_q\right.\right.\nn\\ &&\left.\left.
- c\Omega_{q-n}- c^2\right]+2 n^2\left[10 \omega_q \Omega_{q-n}-8 q\left(-n+q\right)+5 c^2-6 c \omega_q-3 c\Omega_{q-n}\right]\right.\nn\\ &&\left.
-24 n\left(\omega_{n}-\frac{c}{2}\right)\left[-2\left(-n+q\right)\omega_q+q \Omega_{q-n}+2 c q- c n\right]\right]
\label{s3hatbis}
\end{eqnarray}
and
\begin{equation}
\sum_{|i_2\rangle\neq |s.o.\rangle}\frac{\left|\langle i_2|H_{3,BF}|s.o.\rangle\right|^2}{E^{(0)}_{|s.o.\rangle}-E^{(0)}_{|i_2\rangle}}=\hat \CS_4~,
\label{s4hat}
\end{equation}
with
\begin{eqnarray}
\hat \CS_4=&&-\sum_{q=-N}^N\frac{1}{32 \,c \, \omega_n\omega_q\Omega_{q+n}\left(\omega_n+\omega_q+\Omega_{q+n}\right)}\left[18\left(\omega_n-\frac{c}{2}\right)^2\left[2 \omega_q\Omega_{q+n}-2 c \omega_q\right.\right.\nn\\ &&\left.\left.
+c\Omega_{q+n}- c^2\right]+2 n^2\left[10 \omega_q \Omega_{q+n}-8 q\left(n+q\right)+5 c^2+6 c \omega_q+3 c\Omega_{q+n}\right]\right.\nn\\ &&\left.
+24 n\left(\omega_{n}-\frac{c}{2}\right)\left[2\left(n+q\right)\omega_q-q \Omega_{q+n}+2 c q+c n\right]\right]
\label{s4hatbis}
\end{eqnarray}

Using the symmetric prescription for the normal ordering of the quartic Hamiltonian, it is straightforward to derive
\begin{equation}
\langle s.o |H_{4}|s.o.\rangle_{\rm sym} =\frac{1}{2}\left(\CS_5+\CS_6+\CS_7+\CS_8\right)
\end{equation}
and therefore
\begin{equation}
E_{s.o.}^{(2)}=\hat \CS_1+\hat \CS_2+\hat \CS_3+\hat \CS_4+\frac{1}{2}\left(\CS_5+\CS_6+\CS_7+\CS_8\right)=\frac{1}{2}\CS(n)~.
\end{equation}
This result provides additional evidence of the fact that the sum $\CS(n)$, which we shall discuss in detail and evaluate in the next section, appears in the spectrum as a finite size correction to the dispersion relation of the single magnon.


\section{Regularization prescription}
\label{regularization}

In this Section we examine the question of how to regularize the divergent sums appearing in all the results of the previous section. In fact, there is an ongoing discussion in the literature about this issue~~\cite{Krishnan:2008zs, McLoughlin:2008ms, Alday:2008ut,  Gromov:2008fy, Bandres:2009kw,Abbott:2010yb} and hopefully our analysis will contribute in understanding it better. To illustrate our prescription we focus on how to compute the sum~\eqref{esse}.

There are essentially two different prescriptions for how to sum over of the mode numbers.  In the following we present both prescriptions and propose a solution for the discrepancies on the results, which leads to the final expressions \eqref{senergyintro} and \eqref{tenergyintro} that we gave in the Introduction.

As we have seen, the form of the cubic interacting Hamiltonian $H_3$, see Eqs.~\eqref{hc1osc} and \eqref{H3BFoscillators}, naturally implies that if we choose a cutoff $N$ when summing over the light modes, this automatically gives that the heavy modes should have a cutoff equal to $2N$ leading to Eq.~\eqref{esse}. This can be rearranged as
\begin{eqnarray}
\label{esse1}
\CS(n)&=&\frac{1}{2\,c\,R^2\omega_n}\left[8n^2\left(\sum_{q=-2N}^{2N}\frac{1}{\Omega_q}-\sum_{q=-N}^{N}\frac{1}{\omega_q}\right)\right.\cr&+&\left.
\left(\frac{9}{2}c^2-9c\,\omega_n+8n^2\right)\left(\sum_{q=-2N}^{2N}\frac{1}{\Omega_{q}}-\sum_{q=-N}^{N}\frac{1}{\Omega_{q+n}}\right)\right]
\end{eqnarray}
In order to safely remove the cutoff we should manipulate \eqref{esse1} so that all the sums have the same cutoff. We thus get
\begin{equation}
\label{esse2}
\CS(n)=\frac{1}{2\,c\,R^2\omega_n}\sum_{p=-N}^{N}\left[8n^2\left(\frac{1}{\Omega_{2p+1}}-\frac{1}{\Omega_{2p}}\right)+
\left(\frac{9}{2}c^2-9c\,\omega_n+8n^2\right)\left(\frac{1}{\Omega_{2p+1}}+\frac{1}{\Omega_{2p}}-\frac{1}{\Omega_{p+n}}\right)\right]
\end{equation}
Since we know that $\CS(n)$ is actually convergent, we can now send $N\to\infty$ to remove the cutoff. All the sums in \eqref{esse2}  can be computed by standard $\zeta$-function techniques.
To give them a precise (regularized) definition one introduces the function
 \begin{equation}
G(s)=\sum_{p=-\infty}^{\infty}\frac{1}{[(p+a)^2+b^2]^s}
\end{equation}
and considers its analytic continuation for complex $s$.
Then for ${\rm Re}(s)>1/2$ one can write
\begin{equation}\label{Gs}
G(s)=\frac{\sqrt{\pi }}{\Gamma (s)} \left[\frac{\Gamma(s-1/2)}{b^{2s-1}}+4 \pi ^{s-\frac{1}{2}} \sum _{p=1}^{\infty }
   e^{2\pi i p a}\left(\frac{p}{b}\right)^{s-\frac{1}{2}} K_{\frac{1}{2}-s}(2 b p \pi
   )\right]
\end{equation}
where $K_\nu(x)$ is the modified  Bessel function defined, for ${\rm Re}(x)>0$, by its integral representation
\begin{equation}
K_\nu(x)=\frac{1}{2}\int_0^\infty\frac{dv}{v} v^{\nu}e^{-\frac{x}{2}(v+1/v)}
\end{equation}
Taking the limit $s\to 1/2$, which is what interests us, we obtain
\begin{equation}
G(s)\simeq\frac{1}{s-\frac{1}{2}} +2 \log (2)-2 \log
   (b)+4 \sum _{p=1}^{\infty }e^{2\pi i p a} K_0(2 b p \pi )+\mathcal{O}\left(s-\frac{1}{2}\right)
\end{equation}
The pole does not depend on $b$, therefore, as expected,  it cancels in $\CS(n)$. The remaining finite result is
\begin{eqnarray}\label{csn}
\CS(n)=\frac{8 n^2}{\,c\,R^2\omega_n}\sum_{p=1}^{\infty}\left[(-1)^p-1\right]K_0(\pi c p)
\end{eqnarray}
In the limit for large $c$ it is exponentially suppressed
\begin{eqnarray}
\CS(n)&=&\frac{2\sqrt{2} n^2}{R^2 c}\,e^{-\pi c} \left[\frac{8 }{c^{3/2}}-\frac{1}{\pi
   c^{5/2}}+\CO\left(\left(\frac{1}{c}\right)^{7/2}\right)\right]+\CO\left(e^{-2 \pi c}\right)\cr&=&
   \frac{ (\sqrt{2}\pi) ^{3/2} n^2}{J}e^{-\frac{J}{ \sqrt{2\lambda }}} \left[4
   \sqrt{2} \left(\frac{\sqrt{\lambda
   }}{J}\right)^{3/2}-\left(\frac{\sqrt{\lambda}}{J}\right)^{5/2}+\CO\left(\left(\frac{\sqrt{\lambda}}{J}\right)^{7/2}\right)\right]
   +\CO\left(e^{-\frac{\sqrt{2}J}{ \sqrt{\lambda }}}\right)\cr&&
\end{eqnarray}
where we have expressed it also in terms of gauge theory quantities. This is the non analytic, exponentially suppressed, part of the finite size correction that could not be computed in \cite{Astolfi:2008ji}. It appears in all the string states we have considered and can be ascribed to a finite size correction of the dispersion relation \eqref{magnon_disp_rel} arising, in the dual gauge theory, from wrapping interactions (see the recent review on the subject~\cite{Janik:2010kd} and reference therein). For the expansion of $\CS(n)$ for small $c$ (large $\lambda'$) see Appendix~\ref{app:smallc}.

The result \eqref{csn} gives for the coefficient $a_1$ in the strong coupling expansion of the interpolating function $h(\lambda)$ $a_1=0$, see \eqref{dispersioncorrected}. Note that this result is also in agreement with the conjectured exact form of $h(\lambda)$ proposed in Ref. \cite{Leoni:2010tb} from a weak coupling calculation.

The prescription that we consider here is in agreement with the one used in \cite{Gromov:2008fy, Bandres:2009kw} where it was first pointed out that a suitable prescription for summing over the world-sheet frequencies should distinguish between heavy and light modes.
In particular, the authors of~\cite{Gromov:2008fy} proposed a regularization in order to restore agreement among the results obtained in the context of the string world-sheet one-loop analysis~\cite{McLoughlin:2008ms, Alday:2008ut, Krishnan:2008zs, McLoughlin:2008he} and those obtained using the all-loop Bethe ansatz. In \cite{Gromov:2008fy} the authors proposed a regularization of the sums such that the heavy excitations with mode number $q$ should be treated on equal footing as light excitations of mode number $q/2$.
In \cite{Bandres:2009kw} a similar prescription was proposed in order to obtain finite one loop corrections to the energy of circular strings and basically to have consistency between the algebraic curve and supersymmetry. The motivation behind this prescription is essentially that the heavy excitations are not fundamental but rather bound states of two light fundamental modes~\cite{Zarembo:2009au}. This leads to the choice of a different cutoff for the two kinds of excitations. Here we show that the interaction Hamiltonian forces the cutoff on heavy modes to be twice that on light modes.

There is however another proposal for how to regularize the sum~\eqref{esse} which was used
in \cite{McLoughlin:2008ms, Alday:2008ut, Krishnan:2008zs, McLoughlin:2008he} in the context of the semiclassical world-sheet computation of the folded and spinning string in $\ads_4 \times \C P^3$, see also~\cite{Mikhaylov:2010ib}. This prescription was also adopted in the $\ads_5 / \cft_4$ case and it does not distinguish between the world-sheet heavy and light excitations. This implies that in Eq.~\eqref{esse} we simply remove the cutoff by sending $N\to\infty$ to get
\be
\label{firstprescr}
\CS(n)=\frac{4\,n^2}{R^2c\,\omega_n}\sum_{q=-\infty}^\infty\left(\frac{1}{\Omega_q}-\frac{1}{\omega_q}\right)
\ee
Using \eqref{Gs} in the $s\to 1/2$ limit, yields
\be
\label{sumlog}
\CS(n)=- \frac{8 n^2}{\,c\,R^2\omega_n}\left( \log 2- \sum_{p=1}^{\infty}\left[(-1)^p-1\right]K_0(\pi c p)\right)
\ee
This would give $a_1=-\frac{\log 2}{2\pi}$, but, as we have shown, the cutoffs being different, it is more natural to first write all the sums in terms of a single cutoff and then remove it. Note that the second term in \eqref{sumlog} is independent on the regularization, in \eqref{csn} it appears the same term.

%

\section{Normal ordering prescription}
\label{sec:no}

The Hamiltonian used to compute the energy of string states has been derived as a classical object. In order to quantize it we need to replace each field in terms of its mode expansion, namely in terms of annihilation and creation operators. In general in going from classical to quantum expressions there is an ambiguity in the order of writing the operators. The standard procedure, which is also the one used in this Paper, is to completely symmetrize (or antisymmetrize in the case of fermionic fields) the fields and rewrite them as quantum operators. This corresponds to a precise choice of the normal ordering prescription. Here we will show that this prescription appears as a natural consequence of the requirements that the energy of string states has to be finite and that the pp-wave algebra has to be satisfied.

The normal-ordering ambiguity is introduced through appropriate normal-ordering constants~\footnote{Note that in principle one can introduce normal ordering "functions", but in this Paper, without loss of generality, we only consider the case of normal ordering constants.}.
Generically we write for two oscillators
\be
\label{2oscillators}
&& ({a^\dagger}_m \, a_p)_C = (a_p \, {a^\dagger}_m)_C \longrightarrow \alpha \,{a^\dagger}_m \, a_p+(1-\alpha )\, a_p \,{a^\dagger}_m
\cr
&& ({b^\dagger}_m \, b_p)_C \equiv -(b_p \, {b^\dagger}_m)_C \longrightarrow \beta\, {b^\dagger_m}\, b_p+(\beta-1)\, b_p\, {b^\dagger}_m
\ee
where the subscript $C$ refers to the classical object, $a, {a^\dagger}$ are  bosonic annihilation and creation operators and $b, {b^\dagger}$ are the corresponding fermionic quantities. $\alpha\,, \beta$ are normal ordering constants which encode the ambiguity in rearranging the oscillators when one derives operators from classical expressions.

In computing corrections to the energy of string states up to the order $1\over R^2$, we also need to specify the normal ordering prescription of terms cubic and quartic in the number of oscillators. The question of normal ordering for the cubic Hamiltonian, Eqs. \eqref{hc1osc} and \eqref{H3BFoscillators}, it is analogous to the one for the quadratic Hamiltonian, Eqs. \eqref{penspectrum} and \eqref{fermppwave}, since one of the oscillators in the cubic Hamiltonians is always different from the other two.
However, to complete our analysis, we should also consider the normal ordering of terms which are quartic in the number of oscillators.
This can be considered as a generalization of the two-oscillator case \eqref{2oscillators} and the solution will follow similarly: for each family of oscillators the quantum operator can be written as a linear combination of all the possible ways of ordering the annihilation and creation operators, with the constraint that the sum of the normal ordering constants must be equal to 1
\begin{eqnarray}
 (a^\dagger_{-m}a^\dagger_{-p} a_q a_r)_C  & \longrightarrow & \alpha_1\, a^\dagger_{-m} a^\dagger_{-p} a_q a_r
 +  \alpha_2 \,a^\dagger_{-m} a_q a^\dagger_{-p} a_r
+ \alpha_3 \,a^\dagger_{-m} a_q a_r a^\dagger_{-p}\cr
&+& \alpha_4\,a_q a^\dagger_{-m} a_r a^\dagger_{-p}
+ \alpha_5\,a_q a^\dagger_{-m} a^\dagger_{-p} a_r
+ \alpha_6\,a_q a_r a^\dagger_{-m}a^\dagger_{-p}\label{quarticnormalorder}
\end{eqnarray}
where $\sum_{i=1}^6 \alpha_i=1$.
Note that there is a set of six constants $\alpha_i$ for each type of oscillator.

We also make the assumption that families of oscillators having degenerate plane-wave energy have the same normal ordering constant~\footnote{Note that this is also true in flat space.}. Concretely this means that the set of oscillators labeled by  $a^1$ and $a^2$ have the same normal ordering constant $\alpha$, as well as the set of oscillators $\tilde a^1$ and $\tilde a^2$ that have the same normal ordering constant $\tilde \alpha$. The same is true for the four heavy bosons $\hat a^i$, with $i=1, \dots,4$ with the corresponding normal ordering constant $\alpha_u$ and for the four light fermions $d_i$, $i=1,\dots,4$, with normal ordering constant $\alpha_d$. Finally, since the four heavy fermions split in two degenerate families, labeled by the eigenvalues $\pm 1$ of the matrix $i\Gamma_{56}$, cf. section \ref{app:plane_wave}, they have two different normal ordering constants $\alpha_{b1}$ and $\alpha_{b2}$, respectively.

Summarizing, we have 6 normal ordering constants for terms involving two-oscillators
\be
\label{no_constants}
\alpha\,,\, \tilde\alpha\,,\, \alpha_u\,,\, \alpha_d\,,\, \alpha_{b1}\,, \,\alpha_{b2}\,
\ee
and from the terms with four oscillators \eqref{quarticnormalorder} we have 5 independent normal-ordering constants for each type of oscillator.
>From the study of the pp-wave algebra we will obtain constraints on the normal ordering constants.

\subsection{Normal ordering and plane-wave algebra}
\label{sec:pw_algebra}

The bosonic generators of the pp-wave algebra are
\be
L_{ij}\,,~ i\neq j=1,2,3\,,~~~  L_{56},~~~L_{78}, ~~~ H_2\,
\ee
where $L_{ij}=-L_{ji}$ are the rotation generators in the transverse directions of $\ads_4$. Despite the fact that the bosonic quadratic Lagrangian \eqref{CL2} is invariant with respect to rotation involving also the coordinate $u_4$, namely under the rotation $\delta u_i =\epsilon_{ij} u_j$ with $i=1, \dots, 4$, this is not true for the fermionic plane-wave Lagrangian \eqref{finalCL2F}. Indeed, such a transformation involves the combination $\Gamma_{i4}$ which does not leave the fermionic Lagrangian invariant.

$L_{56}$ and $L_{78}$ are the generators of the rotations in the transverse directions in $\C P^3$.  The Lagrangian is not invariant under a rotation involving an arbitrary couple of transverse directions in $\C P^3$, say for example under $L_{57}$, since the Penrose limit we are taking selects two flat directions the 5 and the 7. Finally $H_2$ is the plane-wave Hamiltonian.

Obviously, the full plane-wave supersymmetry algebra includes also the transverse translation currents and the plane-wave supercharges (fermionic generators). However, since we want to investigate the effect of the normal ordering on these generators, we need only to consider generators which are quadratic in the number of fields. The translation currents, apart from $H_2$, are linear in the fields and of course do not suffer from ordering ambiguities. The supercharges are instead fermionic operators made of a product of a fermionic and a bosonic oscillator and consequently they do not have any ordering problem at the plane-wave level. Hence, for our discussion we explicitly construct the subsector of the full plane-wave symmetry which is needed for the purpose of understanding normal ordering issues, following~\cite{Metsaev:2001bj}, where the plane-wave algebra for the $\ads_5 \times\sphere^5$ superstring was worked out.

\subsubsection*{Angular momenta}

Let us consider an infinitesimal rotation in one $\grSU(2)$ subsector of $\C P^3$, for example
\begin{equation}
\delta x_1 =\epsilon y_1\,, \qquad \delta y_1 =-\epsilon x_1 \,, \qquad \delta \theta =\epsilon\Gamma_{56}\theta\,
\end{equation}
where $\epsilon$ is an infinitesimal parameter.
After imposing the light-cone gauge condition~\eqref{lc_gf}, under such a rotation the plane-wave Lagrangian (see Appendix \ref{app:plane_wave})  transforms as
\begin{equation}
\delta \mathcal{L}_2=\frac{\epsilon c}{8}\left(\dot y_1 y_1-\dot x_1 x_1\right)=\frac{\epsilon c}{16} \frac{d}{d\tau}\left(y_1^2-x_1^2\right)\,
\end{equation}
i.e. the variation of the Lagrangian is a total derivative. Let us now consider the generator of rotations in the 5\,6 plane, relative to the directions $x^1$ and $y^1$. This corresponds to $\tilde{L}_{56}$ given by
\begin{eqnarray}
\label{Ltilde_56}
 \tilde L_{56}  =  \int_0^{2\pi}  \frac{d\sigma}{2\pi} \left(p_{x1}y_1-p_{y1} x_1-\frac{i c}{2}\bar\theta\Gamma_{+56}\theta\right)
\end{eqnarray}
We see that the commutator of $\tilde{L}_{56}$ with the plane-wave Hamiltonian is a total derivative.
This implies that the conserved quantity is given by
$L_{56}=\tilde L_{56}-\frac{c}{16}\left(y_1^2-x_1^2\right)$
which in fact commutes with the Hamiltonian.

In order to write down explicitly the conserved charges, we have to consider the eigenstates of $\Gamma_{56}$ which are labelled as follows
\be
\label{Gamma56_eigenstates}
&& \Gamma_{56}\, d_{1,2}= i d_{1,2}\,,\qquad \Gamma_{56}\, d_{3,4}=-i d_{3,4}\,
~~~~
\Gamma_{56}\, b_{1,2}= i b_{1,2}\,,\qquad \Gamma_{56}\, b_{3,4}=-i b_{3,4}\,
\ee
Using this notation, one can write the angular momentum $L_{56}$ in terms of the oscillators
\begin{eqnarray} \label{L56oscillators}
L_{56}  & = &
\sum_{m\in \Z }\big(a^\dagger_m a_m-\tilde a^\dagger_m \tilde a_m-b^\dagger_{1,m}b_{1,m}-b^\dagger_{2,m}b_{2,m}+b^\dagger_{3,m}b_{3,m}+b^\dagger_{4,m}b_{4,m}
\cr
&-& d^\dagger_{1,m}d_{1,m}-d^\dagger_{2,m}d_{2,m}+d^\dagger_{3,m}d_{3,m}+d^\dagger_{4,m}d_{4,m}\big)
+ \sum_{m \in \Z} \left(\tilde \alpha-\alpha-2\alpha_{b1}-2\alpha_{b2}\right)~~
\end{eqnarray}
Similarly one computes the generator of rotations in the other $\grSU(2)$ sector, $L_{78}$. We obtain
\begin{eqnarray}
\label{L78oscillators}
 L_{78}  &= & \sum_{m}\big(a^\dagger_{2,m} a_{2,m}-\tilde a^\dagger_{2,m} \tilde a_{2,m}-b^\dagger_{1,m}b_{1,m}-b^\dagger_{2,m}b_{2,m}+b^\dagger_{3,m}b_{3,m}+b^\dagger_{4,m}b_{4,m}
 \cr
  &+&
d^\dagger_{1,m}d_{1,m}+d^\dagger_{2,m}d_{2,m}-d^\dagger_{3,m}d_{3,m}-d^\dagger_{4,m}d_{4,m}\big)
+ \sum_m\left(\tilde \alpha-\alpha-2\alpha_{b1}-2\alpha_{b2}\right)\,
\end{eqnarray}

Let us now consider infinitesimal rotations in the transverse directions of $\ads_4$ and their effect on the fermionic coordinates, i.e.
\begin{equation}
\label{urot}
\delta u_i =\epsilon_{ij} u_j\,,\qquad \delta \theta = \epsilon_{ij}\Gamma_{ij}\theta\,, \qquad i,j=1,2,3\,
\end{equation}
with $\epsilon_{ij}=-\epsilon_{ji}$ and $i=1,2,3$.
Such transformations are symmetries of the plane-wave Lagrangian, namely $\delta \mathcal{L}_{2}=0$, where the subscript 2 refers to the number of fields.

For the rotation \eqref{urot} the corresponding angular momenta are
\begin{equation} \label{angmomentaui}
L_{ij}=\int_0^{2\pi} {d\sigma \over 2\pi} \left(u_i \dot u_j-\dot u_i u_j-\frac{i c}{2}\bar\theta \Gamma_{+ij}\theta\right)\,, \qquad  i,j=1,2,3\,
\end{equation}
There are three independent generators associated to this rotational invariance: $L_{12}$, $L_{13}$ and $L_{23}$. They obey the standard commutation relations thus we can diagonalize only one of them simultaneously with $L_{56}$ and $L_{78}$.
For example we can focus on $L_{12}$ which is given by
\begin{equation}
L_{12}=\frac{1}{2\pi}\int d\sigma\left(u_1 p_{u2}-u_2 p_{u1}-\frac{i c}{2}\bar\theta \Gamma_{+12}\theta\right)\,
\end{equation}
We have
\begin{eqnarray} \label{L12}
 L_{12} & = & \sum_m \big(i a^\dagger_{u1,n} a_{u2,n}- i a_{u1,n} a^\dagger_{u2,n}-b^\dagger_{1,m}b_{1,m}+b^\dagger_{2,m}b_{2,m}+b^\dagger_{3,m}b_{3,m}-b^\dagger_{4,m}b_{4,m}\cr
& - & d^\dagger_{1,m}d_{1,m}+d^\dagger_{2,m}d_{2,m}+d^\dagger_{3,m}d_{3,m}-d^\dagger_{4,m}d_{4,m}\big)
\end{eqnarray}
Finally we should require that the expectation value of the plane-wave symmetry generators $L_{56}$, $L_{78}$ and $L_{12}$ on the vacuum must vanish. This requirement leads to the following constraint on the normal ordering constants
\begin{equation} \label{constrangmom}
\tilde \alpha-\alpha-2\alpha_{b1}-2\alpha_{b2}=0\,
\end{equation}
%

\subsubsection*{Normal ordering in the Hamiltonian}

Another interesting generator is the plane-wave gauge fixed Hamiltonian, which in terms of  oscillators is given by
\begin{eqnarray} \label{Hpp}
c H_2 & = &
\sum_{m\in \Z} \Big\{ \sum_{a=1}^2\left[ M^a_m \left(\omega_m-\frac{c}{2}\right)+N^a_n\left(\omega_m+\frac{c}{2}\right)\right]+\sum_{i=1}^4 \hat N^i_n \Omega_m
\cr
&+ & \sum_{f=1}^4 \omega_m F_m^{(f)}+\sum_{f=5}^6\left(\Omega_m+\frac{c}{2}\right)F_m^{(f)}+\sum_{f =7}^8\left(\Omega_m-\frac{c}{2}\right)F_m^{(f)}\cr
&+& \sum_m 2\omega_m\left(2\alpha_{d}-\alpha-\tilde\alpha\right)+c\left(\alpha-\tilde\alpha+\alpha_{b1}-\alpha_{b2}\right)+2\Omega_m\left(\alpha_{b1}+\alpha_{b2}
-2\alpha_{u}\right)\Big\}\,
\end{eqnarray}
Imposing that the vacuum expectation value of the plane-wave Hamiltonian is not affected by the normal ordering constants gives the following constraints
\be
\label{ppwavehamconstr}
2\alpha_{d}-\alpha-\tilde\alpha=0~~,~~~\alpha-\tilde\alpha+\alpha_{b1}-\alpha_{b2}=0
~~,~~~
\alpha_{b1}+\alpha_{b2}
-2\alpha_{u}=0
\ee

The cubic Hamiltonian does not impose further constraints, since the normal ordering constants cancel, once we impose the requirements that oscillators with the same pp-wave energy have the same normal ordering constants.

\subsection{Cancelation of divergences}
We now want to consider the constraints on the normal ordering constants coming from requiring that the energy corrections to the string states considered in Section~\ref{sec:string_finitesize_en} are free of divergences. Eqs.\eqref{H4BBsym} and \eqref{H4BFsym} can be recalculated keeping into account the general normal ordering prescription we are considering in this section to get
\begin{eqnarray}
\label{samesu2spectrum1}
&&\langle s| H_{4,B}
|s\rangle =
-\frac{2\left[\left(\omega_n-c\right)\left(4n^2-c^2\right)-c^2
\omega_n\right]}{ c^3 \omega_n}\cr
&+& \sum_{q=-N}^N\frac{2}{c^3 \omega_n \omega_q}\bigg\{ \left(1-\tilde \alpha\right)\bigg[-2 q^2 n^2-2\left(\omega_q+\frac{c}{2}\right)^2\left(\omega_n-\frac{c}{2}\right)^2
- 4 c^2\left(\omega_q+\frac{c}{2}\right)\left(\omega_n-\frac{c}{2}\right)\bigg]
\cr
& -&\frac{1}{2} \left(\alpha_2+2 \alpha_3+3 \alpha_4+2 \alpha_5+4 \alpha_6\right)
\bigg[q^2 n^2 + \left(\omega_q-\frac{c}{2}\right)^2\left(\omega_n-\frac{c}{2}\right)^2
- 2 c^2\left(\omega_q-\frac{c}{2}\right)\left(\omega_n-\frac{c}{2}\right)\bigg]
\cr
&- & \left[n^2+\left(\omega_n-\frac{c}{2}\right)^2\right] \bigg[ \left(1-\alpha\right)\left( q^2+\left(\omega_q-\frac{c}{2}\right)^2\right)
-\frac{2 c^2\omega_q\left(1-\alpha_{u}\right)}{\Omega_q}
\cr &+ &\left(1-\tilde \alpha\right)\left( q^2+\left(\omega_q+\frac{c}{2}\right)^2\right)    \bigg]
\cr
&+ &\frac{2 c^2\omega_q\left(1-\alpha_{u}\right)}{\Omega_q }\left(\omega_n-\frac{c}{2}\right)^2-\frac{4\omega_q\left(1-\alpha_{u}\right)}{ \Omega_q }\left(2 q^2+c^2\right)
\left[n^2+\left(\omega_n-\frac{c}{2}\right)^2\right]\bigg\}
\end{eqnarray}
\begin{eqnarray}
\label{H4BF_meanvaluebis}
&& \langle s | H_{4,BF} | s\rangle =-\frac{1}{2\,c\, \omega_n}\sum_{q=-\infty}^\infty\bigg\{9\left(\omega_n-\frac{c}{2}\right)^2\left[\frac{\left(\alpha_{b1}+\alpha_{b2}-2\right)}{\Omega_q}+\frac{\left(1-\alpha_{d}\right)}{\omega_q}\right]\cr
&+&11 n^2\left[ \frac{\left(\alpha_{b1}+\alpha_{b2}-2\right)}{\Omega_q}
+\frac{\left(\alpha_{d}-1\right)}{\omega_q}\right]
+
\frac{16\, q^2\left[n^2+\left(\omega_n-\frac{c}{2}\right)^2\right]}{c^2}\left[ \frac{\left(\alpha_{b1}+\alpha_{b2}-2\right)}{\Omega_q}
+2\frac{\left(\alpha_{d}-1\right)}{\omega_q}\right]\bigg\}\cr &&
\end{eqnarray}
Putting this results together with the contributions coming from the cubic Hamiltonian Eq.\eqref{cubicspectrum} and requiring cancelation of divergences we get the following constraints on the normal ordering constants
\begin{eqnarray}
&&\frac{1}{4}(\alpha_2+2\alpha_3+3\alpha_4+2\alpha_5+4\alpha_6)+2\left(\alpha_{b_1}+\alpha_{b_2}+2\alpha_d-2\alpha_u-\tilde \alpha\right)-\alpha_a=1
\nn\\
&&
\frac{1}{2}\left(\alpha_2+2\alpha_3+3\alpha_4+2\alpha_5+4\alpha_6\right)+\alpha_a+\tilde\alpha=2
\nn\\
&&
\frac{1}{4}(\alpha_2+2\alpha_3+3\alpha_4+2\alpha_5+4\alpha_6)+2\tilde \alpha-\alpha_a=1
\nn\\
&&
\frac{5}{2}\left(\alpha_2+2\alpha_3+3\alpha_4+2\alpha_5+4\alpha_6\right)+2\left(\alpha_a+13\alpha_d-\alpha_{b_1}-\alpha_{b_2}-4\tilde \alpha-16\alpha_u\right)=-3
\nn\\
&&
\frac{1}{2}\left(\alpha_2+2\alpha_3+3\alpha_4+2\alpha_5+4\alpha_6\right)+2\left(2\alpha_{b_1}+2\alpha_{b_2}+12 \alpha_u-\alpha_a-3\alpha_d-2\tilde \alpha\right)=11
\label{divcanc}
\end{eqnarray}
Note that the normal ordering constants $\alpha_i$, with $i=1,\dots, 6$, for the term with 4 oscillators of the same kind always appear in the same combination. Solving simultaneously Eqs.~\eqref{constrangmom}, \eqref{ppwavehamconstr} and~\eqref{divcanc}, we find that there is a unique solution given by
\begin{equation} \label{solutionconstants}
\alpha_a=\tilde \alpha =\alpha_u=\alpha_d=\alpha_{b_1}=\alpha_{b_2}=\frac{1}{2}~~~,~~~\alpha_2+2\alpha_3+3\alpha_4+2\alpha_5+4\alpha_6=2
\end{equation}
If we moreover require, for the normal ordering prescription of the terms with 4 oscillators of the same kind, that the $\alpha_i$'s are all equal, we get that they have to be $\alpha_i=\frac{1}{6}$ for $i=1,\dots, 6$.
This is precisely the symmetric prescription that we used in Section~\ref{sec:string_finitesize_en}.

\section{Finite size corrections for states outside the $\grSU(2)\times \grSU(2)$ subsector}
\label{sec:outside_SU2xSU2}

The $\grSU(2)\times \grSU(2)$ subsector has been our privileged testing ground. It naturally decouples in the plane-wave limit and furthermore the interacting Hamiltonian up to order ${1\over R^2}$ is diagonal, in the sense that the states $|s\rangle$ and $|t\rangle$ $\in\grSU(2)\times \grSU(2)$, do not get mixed up at this order.
This is a very special feature of the closed $\grSU(2)\times \grSU(2)$ subsector, indeed in general two-magnon states are degenerate and they are mixed by the perturbation Hamiltonian. Now we want to take some steps toward a more general description of the near-plane wave spectrum. We defer to a forthcoming paper~\cite{astolfifinal} for a complete study of the near plane wave spectrum of type IIA string states in $\ads_4\times \C P^3$.

We want here to compute the energy up to order $1/R^2$ of two-impurity states which are still in $\C P^3$ but not in $\grSU(2)\times \grSU(2)$ and, most importantly, which are not degenerate. Such states are
\be
\label{stilde}
&& |\tilde s\rangle = (\tilde a_n^1)^\dagger (\tilde a_{-n}^1)^\dagger|0\rangle\,,~~~~~
 |\tilde t\rangle =\frac{1}{2}\left( (\tilde a^1_n)^\dagger (\tilde a^2_{-n})^\dagger+(\tilde a^1_{-n})^\dagger (\tilde a^2_{n})^\dagger\right)|0\rangle
\ee

The spectrum is given by
\begin{equation}\label{c}
E_{\tilde s,\tilde t}^{(2)}=\sum_{|i\rangle}\frac{\left|\langle i|H_{3}|\tilde s,\tilde t\rangle\right|^2}{E^{(0)}_{|\tilde s\rangle,|\tilde t\rangle}-E^{(0)}_{|i\rangle}}+\langle \tilde s, \tilde t |H_{4}|\tilde s,\tilde t\rangle
\end{equation}
where $|i\rangle$ is an intermediate state with zeroth order energy
$E^{(0)}_{|i\rangle}$. The computation proceeds exactly as in Section~\ref{sec:string_finitesize_en}, so here we report only the results.

Consider first the state $|\tilde s\rangle$. The intermediate states contributing to the first term in~\eqref{c} are
\begin{equation}| i_1 \rangle =(a^4_{-p-q})^\dagger
(\tilde a^1_{p})^\dagger(\tilde a^1_{q})^\dagger|0\rangle~~~~\hbox{and}~~~~ | i_2 \rangle =(a^4_{-p-q-r-s})^\dagger
(\tilde a^1_{p})^\dagger(\tilde a^1_{q})^\dagger(\tilde a^1_{r})^\dagger(a^1_{s})^\dagger| 0 \rangle
\end{equation}
for $H_{3,B}$ and
\begin{equation}
| i_3\rangle = (\tilde a^1_r)^\dagger d^\dagger_{q,\alpha} b^\dagger_{-r-q, \beta}| 0\rangle~~~~\hbox{and}~~~~ | i_4\rangle = (\tilde a^1_r)^\dagger (\tilde a^1_t)^\dagger (\tilde a^1_u)^\dagger d^\dagger_{q,\alpha} b^\dagger_{-u-t-r-q, \beta}| 0\rangle
\end{equation}
for $H_{3,BF}$. Proceeding as in Section~\ref{sec:string_finitesize_en}, we find for the energy
\begin{equation} \label{stildeenergy}
E_{\tilde{s}}^{(2)}=
-\frac{8\,n^2\left[\left(\omega_n+\frac{c}{2}\right)^2-\frac{c^2}{2}\right]}{R^2c^3\omega_n^2}
+ \tilde\CS(n)~,
\end{equation}
where
\begin{eqnarray} \label{essetilde}
\tilde\CS(n)&=&\frac{1}{2\,c\,R^2\omega_n}\left[8n^2\left(\sum_{q=-2N}^{2N}\frac{1}{\Omega_q}-\sum_{q=-N}^{N}\frac{1}{\omega_q}\right)\right.\cr&+&\left.
\left(-\frac{9}{2}c^2+9c\,\omega_n+8n^2\right)\left(\sum_{q=-2N}^{2N}\frac{1}{\Omega_{q}}-\sum_{q=-N}^{N}\frac{1}{\Omega_{q+n}}\right)\right]
\end{eqnarray}
Similarly one can compute the energy at order $1/R^2$ of the state $|\tilde t\rangle$. It reads
\begin{equation} \label{ttildeenergy}
E_{\tilde{t}}^{(2)}=-\frac{4\,n^2\left(\omega_n+\frac{c}{2}\right)^2}{R^2c^3\omega_n^2}
+ \tilde\CS(n)~,
\end{equation}
where for the computation of the second term in Eq.~\eqref{c}, we used the same normal ordering prescription adopted in Section~\ref{sec:string_finitesize_en}. We see that, as it should be, also for states outside the $\grSU(2)\times \grSU(2)$ sector the energy is free of divergences.
Carefully removing the cutoff by first requiring that all the sums have the same cutoff $N$ we arrive at
\begin{equation}
\tilde\CS(n)=\CS(n)=\frac{8 n^2}{\,c\,R^2\omega_n}\sum_{p=1}^{\infty}\left[(-1)^p-1\right]K_0(\pi c p)
\end{equation}
and therefore the spectrum of the states $|\tilde s \rangle$ and $| \tilde t \rangle$ receives, in addition to the first terms in Eqs. \eqref{stildeenergy} \eqref{ttildeenergy}, the same type of exponentially suppressed finite size corrections of the states in the $\grSU(2)\times \grSU(2)$ sector.


\section{Conclusions}
\label{sec:conclusions}

The results of this Paper show that the Hamiltonian of type IIA superstring on $\ads_4 \times \C P^3$ that we derived in \cite{Astolfi:2009qh} is perfectly consistent and provides finite results for the finite size corrections to the energies of string states, which can be explicitly computed. Moreover the form of the cubic Hamiltonian derived in \cite{Astolfi:2009qh} provides an argument for choosing the appropriate prescription to define the divergent sums appearing in the calculations. The results we obtain show that the strong-weak coupling interpolating function $h(\lambda)$, entering the magnon dispersion relation, does not receive a one-loop correction, in agreement with the algebraic curve spectrum. Therefore the finite size corrections to the energy of strings states in the $\grSU(2)\times \grSU(2)$ are precisely those we computed in \cite{Astolfi:2008ji} plus an exponentially suppressed correction which we explicitly compute and which, from the gauge theory side, s!
 hould arise from wrapping interactions. The leading contributions that have an expansion in integer powers of $\lambda'=\frac{\lambda}{J^2}$ were already shown to be in agreement with the corresponding terms coming from the solutions of the Bethe ansatz equations. It would be extremely interesting to derive also the exponential corrections from the Bethe equations, these should be generated by virtual particles circulating around a circle of finite radius and should be encoded by the so-called L\"uscher corrections.

In \cite{Callan:2003xr} a complete analysis of the spectrum of two oscillator states was performed for type IIB superstring on $\ads_5 \times \sphere^5$. It is clear that a similar study is now at hand also for type IIA superstring on $\ads_4 \times \C P^3$. Since Refs. \cite{Callan:2003xr} provided a milestone contribution to the understanding of integrability of the $\ads_5/\cft_4$ duality, the corresponding analysis for the $\ads_4/\cft_3$ duality would be of great interest~\cite{astolfifinal}.

In this Paper we have also started the study of the algebra of generators in the pp-wave limit. A complete study of the algebra of symmetry generators of this theory was beyond the scope of this Paper, but it could certainly be continued along the line of what was done in~\cite{Arutyunov:2006ak} for type IIB superstring on $\ads_5 \times \sphere^5$.


\acknowledgments

We would like to thank M. C. Abbott, C. Kristjansen, T. McLoughlin, K. Zoubos  for interesting discussions. We are very grateful to K. Zarembo for many enlightening comments and discussions. We also thank A. Zayakin for a careful reading of the manuscript.

\begin{appendix}

\section{Geometrical Set-up}
\label{app:geometrical_setup}


The $\ads_4\times \C P^3$ background has the metric
\begin{equation}
\label{adscp} ds^2 = \frac{R^2}{4} \left( - \cosh^2 \rho dt^2 +
d\rho^2 + \sinh^2 \rho d\hat{\Omega}^2_2 \right) + R^2 ds_{\C P^3}^2
\end{equation}
where the $\C P^3$ metric is
\begin{equation}
\label{cp3} ds_{\C P^3}^2 = d\theta^2 + \frac{\cos^2 \theta}{4}
d\Omega_2^2 + \frac{\sin^2 \theta}{4} d{\Omega_2'}^2 +  4 \cos^2
\theta \sin^2 \theta ( d \delta + \omega )^2
\end{equation}
with
\begin{equation}
\omega = \frac{1}{4} \sin \theta_1 d\varphi_1 + \frac{1}{4} \sin
\theta_2 d\varphi_2
\end{equation}
Here the curvature radius $R$ is given by~\footnote{It is important to point out that the relation between the curvature radius $R$ and the string tension, or the 't Hooft coupling,
can receive quantum corrections due to the fact that the $\ads_4 \times\C P^3$ background is not maximally supersymmetric~\cite{McLoughlin:2008he, Bergman:2009zh}. Such corrections would however affect the interactions only at higher orders.
}
\begin{equation}
\label{radius}
R^4 = 32 \pi^2 \lambda l_s^4
\end{equation}
Furthermore, the $\ads_4\times \C P^3$ background has a constant
dilaton with the string coupling given by
\begin{equation}
g_s = \Big( \frac{32\pi^2 \lambda}{k^4} \Big)^{\frac{1}{4}}
\end{equation}
and it has a two-form and a four-form Ramond Ramond flux which are given by
\begin{equation}
\frac{1}{R} F_{(2)} = - \cos \psi d\psi \wedge ( d\delta + \omega) +
\frac{1-\sin \psi}{4} \cos \theta_1 d\theta_1 \wedge d\varphi_1 -
\frac{1+\sin \psi}{4} \cos \theta_2 d\theta_2 \wedge d\varphi_2
\end{equation}
\begin{equation}
\frac{1}{R^3} F_{(4)} = \frac{3}{8} \epsilon_{\ads_4} = \frac{3}{8}
\cosh \rho \sinh^2 \rho dt \wedge d\rho \wedge d\hat{\Omega}_2
\end{equation}
For our purposes it is
convenient to make the coordinate change
\begin{equation}
\psi = 2\theta - \frac{\pi}{2}
\end{equation}
such that the $\C P^3$ metric \eqref{cp3} takes the form
\begin{equation}
\label{cp32} ds_{\C P^3}^2 = \frac{1}{4} d\psi^2 + \frac{1-\sin
\psi}{8} d\Omega_2^2 + \frac{1+\sin \psi}{8} d{\Omega_2'}^2 + \cos^2
\psi ( d \delta + \omega )^2
\end{equation}

The $\grSU(2)\times \grSU(2)$ sector corresponds to the two two-spheres in
the $\C P^3$ metric \eqref{cp32}, parameterized as
\begin{equation}
d\Omega_2^2 = d\theta_1^2 + \cos^2 \theta_1 d\varphi_1^2 \spa
{d\Omega_2'}^2 = d\theta_2^2 + \cos^2 \theta_2 d\varphi_2^2
\end{equation}
On the string theory side, the $\grSU(2)\times \grSU(2)$ symmetry of the
two two-spheres is a subgroup of the $\grSU(4)$ symmetry of $\C P^3$.
We can take the three independent Cartan generators for the $\grSU(4)$
symmetry to be
\begin{equation}
S_z^{(1)} = - i \partial_{\varphi_1} \spa S_z^{(2)} = - i
\partial_{\varphi_2} \spa J = - \frac{i}{2} \partial_\delta
\end{equation}
where $S_z^{(i)}$ are the Cartan generators of the two two-spheres.

\section{$\grSU(2)\times \grSU(2)$ Penrose limit of $\ads_4\times \C P^3$}
\label{app:penrose}

Consider the $\ads_4\times \C P^3$ metric given by \eqref{adscp} and
\eqref{cp32}. We make the coordinate transformation
\begin{equation}
t' = t \spa \chi = \delta - \frac{1}{2} t
\end{equation}
This gives the following metric for $\ads_4\times \C P^3$
\begin{align}
\label{adscp2} ds^2 = & - \frac{R^2}{4} {dt'}^2 ( \sin^2 \psi  +
\sinh^2 \rho
) + \frac{R^2}{4}( d\rho^2 + \sinh^2 \rho d\hat{\Omega}_2^2 ) \nn \\
& + R^2 \left[ \frac{d\psi^2}{4} + \frac{1-\sin \psi}{8} d\Omega_2^2
+ \frac{1+\sin \psi}{8} d{\Omega_2'}^2 + \cos^2 \psi ( dt' + d \chi
+ \omega )( d \chi + \omega )\right]
\end{align}
We have that
\begin{equation}
\label{newch} E \equiv \Delta - J = i \partial_{t'} \spa 2 J = - i
\partial_\chi
\end{equation}

Define the coordinates
\begin{equation}
v = R^2 \chi \spa x_1 = R \varphi_1 \spa y_1 = R \theta_1 \spa x_2 =
R \varphi_2 \spa y_2 = R \theta_2 \spa u_4 = \frac{R}{2} \psi
\end{equation}
We furthermore define $u_1$, $u_2$ and $u_3$ by the relations
\begin{equation}
\frac{R}{2} \sinh \rho = \frac{u}{1 - \frac{u^2}{R^2}} \spa
\frac{R^2}{4} ( d\rho^2 + \sinh^2 \rho d\hat{\Omega}_2^2 ) =
\frac{\sum_{i=1}^3 du_i^2}{(1-\frac{u^2}{R^2}  )^2} \spa u^2 =
\sum_{i=1}^3 u_i^2
\end{equation}
Written explicitly, the metric \eqref{adscp2} in these coordinates
becomes
\begin{align}
\label{adscp3} &ds^2 = -  {dt'}^2 \left( \frac{R^2}{4} \sin^2
\frac{2u_4}{R} + \frac{u^2}{(1 - \frac{u^2}{R^2})^2} \right) +
\frac{\sum_{i=1}^3 du_i^2}{(1-\frac{u^2}{R^2}  )^2} + du_4^2\nn \\
&
+\frac{1}{8}\left(\cos\frac{u_4}{R}-\sin\frac{u_4}{R}\right)^2\left(
dy_1^2+\cos^2\frac{y_1}{R}d
x_1^2\right)+\frac{1}{8}\left(\cos\frac{u_4}{R}+\sin\frac{u_4}{R}\right)^2\left(
dy_2^2+\cos^2\frac{y_2}{R}d x_2^2\right)
\cr&+R^2\cos^2\frac{2u_4}{R} \left[ dt' + \frac{dv}{R^2}
+\frac{1}{4}\left(\sin \frac{y_1}{R} \frac{d x_1}{R} + \sin
\frac{y_2}{R} \frac{d x_2}{R}\right)\right] \left[\frac{dv}{R^2}
+\frac{1}{4}\left(\sin \frac{y_1}{R} \frac{d x_1}{R} + \sin
\frac{y_2}{R} \frac{d x_2}{R}\right)\right]
\end{align}
a very convenient form to expand around $R\to\infty$.

The $\grSU(2)\times \grSU(2)$ Penrose limit $R \rightarrow \infty$ of
\cite{Grignani:2008is} gives now the pp-wave metric%
\footnote{See \cite{Bertolini:2002nr} for the analogous Penrose
limit for the $\grSU(2)$ sector of $\ads_5\times S^5$.}
\begin{equation}
\label{ppwavemetric} ds^2 =  dv dt'  + \sum_{i=1}^4 ( du_i^2 - u_i^2
{dt'}^2 ) + \frac{1}{8} \sum_{i=1}^2 ( dx_i^2 + dy_i^2 + 2 dt' y_i
dx_i )
\end{equation}
The light-cone coordinates in this metric are $t'$ and $v$. The two-form and four-form
Ramond-Ramond fluxes in the limit are
\begin{equation}
\label{ppwaverrfluxes} F_{(2)} = dt' du_4 \spa F_{(4)} = 3 dt' du_1
du_2 du_3
\end{equation}
This is a pp-wave background with 24 supersymmetries first found in
\cite{Sugiyama:2002tf,Hyun:2002wu}. See
\cite{Nishioka:2008gz,Gaiotto:2008cg} for other Penrose limits of
the $\ads_4\times \C P^3$ background giving the pp-wave background
\eqref{ppwavemetric}-\eqref{ppwaverrfluxes}.

We see from \eqref{newch} that
\begin{equation}
\frac{2 J}{R^2} = - i \partial_v
\end{equation}
Thus, the Penrose limit on the gauge theory side is the following
limit
\begin{equation}
\label{penrlim} \lambda, J \rightarrow \infty \ \ \mbox{with}\ \
\lambda' \equiv \frac{\lambda}{J^2} \ \mbox{fixed} \spa \Delta-J \
\mbox{fixed}
\end{equation}


\section{Plane-wave Lagrangian}
\label{app:plane_wave}

The type IIA GS Lagrangian on $\ads_4\times \C P^3$ was completely worked out in~\cite{Gomis:2008jt,Grassi:2009yj}, based on the supercoset construction of \cite{Metsaev:1998it,Kallosh:1998zx}.
\begin{equation}
\label{generallagr}
\CL = - \frac{1}{2} h^{AB} \eta_{ab} L^a_A L^b_B - 2i \varepsilon^{AB}
\int_0^1 ds L(s)^a_{A} ( \bar{\theta} \Gamma_a \Gamma_{11} )_\alpha
L(s)^\alpha_{B}
\end{equation}
The world-sheet metric is defined as $s_{AB}$ with the world-sheet indices $A,B=0,1$. Then we
define $h^{AB} = \sqrt{|\det s |} s^{AB}$ such that $\det h = -1$. We
furthermore define the epsilon symbol $\varepsilon^{AB}$ such that
$\varepsilon^{01} = \varepsilon_{01} = 1$.
The generalized Maurer-Cartan forms are
\begin{equation}
L(s)_{A}^a = E(s)^a_\mu \partial_A X^\mu  + E(s)_\alpha^a \partial_A
\theta^\alpha  \spa L(s)^\alpha_A = E(s)^\alpha_\mu \partial_A X^\mu
+ E(s)_\beta^\alpha \partial_A \theta^\beta
\end{equation}
with $0 \leq s \leq 1$, $a, b$ the target-space flat indices and $\alpha,\beta$ the spinorial index.

In order to select the 24 supersymmetric fermionic d.o.f. corresponding to the unbroken supersymmetries, we introduce a projector $P$ defined as~\cite{Astolfi:2009qh}
\begin{equation} \label{PP}
P=\frac{3-J}{4}
\end{equation}
with
\begin{equation} \label{J}
J = \Gamma_{0123} \Gamma_{11} ( - \Gamma_{49} -
\Gamma_{56} + \Gamma_{78} ) = \Gamma_{5678} - \Gamma_{49}
(\Gamma_{56} - \Gamma_{78})
\end{equation}
Here we assume that $\theta$ obeys $P \theta=\theta$.
The $\kappa$-symmetry condition that one has to fix for a string entirely moving on $\C P^3$ is~\cite{Astolfi:2009qh}
\be
\label{k_symm_gauge}
(\mathcal P_++\mathcal P_-)\theta=\theta\,,
\ee
where the projectors $\mathcal P_{\pm} $ commute with $P$~\footnote{Notice that $\Gamma_-$ does not commute with $P$. This is why in this case, as opposed to a string moving on $\rm {AdS}_5 \times S^5$ it is not consistent to impose $\Gamma_- \theta=0$ in order to select all the supersymmetric fermions.}
and are defined as
\begin{equation}
\label{pppm}
\begin{array}{c}
 \ds \CP_+ = \frac{I+\Gamma_{5678}}{2}
\frac{I+\Gamma_{4956}}{2}
\spa \CP_- = \frac{I-\Gamma_{5678}}{2}
\frac{I-\Gamma_{09}}{2} \\[3mm]
\end{array}
\end{equation}
and the relation among $P$ and $\CP_{\pm}$ is
\be
&&
P = \CP_+ + \CP_- + \CP_-' \spa I = \CP_+ + \CP_- + \CP_+' + \CP_-'
\\ \nn
&& \CP_+' = \frac{I+\Gamma_{5678}}{2}
\frac{I-\Gamma_{4956}}{2} \spa \CP_-' = \frac{I-\Gamma_{5678}}{2}
\frac{I+\Gamma_{09}}{2}
\ee

Furthermore, $L^a_A = L(s=1)^a_A$ and $L^\alpha_A = L(s=1)^\alpha_A$, and they are constructed from the supervielbeins
\begin{equation}
\label{superviel1}
E(s)^a = e^a +  4 i \bar{\theta} \Gamma^a \frac{\sinh^2 (
\frac{s}{2} \CM )}{\CM^2} D \theta \spa E(s)^\alpha = \left(
\frac{\sinh s \CM }{\CM} D \theta \right)^\alpha
\end{equation}

The covariant derivative is
\begin{equation}
\label{covdertheta}
D \theta = P ( d - \frac{1}{R} \Gamma_{0123} \Gamma_a e^a +
\frac{1}{4} \omega^{ab} \Gamma_{ab} ) \theta
\end{equation}
The two-fermion matrix $\CM^2$ can be found in terms of the structure constants of the generators of $OSp(6|2,2)$~\cite{Astolfi:2009qh}. Schematically we write
\begin{equation}
(\CM^2)^\alpha_\beta = - \theta^\gamma \tilde{f}^\alpha_{\gamma i} \theta^\delta \hat{f}^i_{\delta \beta}
\end{equation}
Finally, the Virasoro constraints are
\begin{equation}
\label{generalvir}
S_{AB} = \frac{1}{2} h_{AB} h^{CD} S_{CD} \spa S_{AB} \equiv
\eta_{ab} L^a_A L^b_B
\end{equation}

We are now briefly describing the pp-wave Lagrangian $\CL_2$ that one obtains in the $R \rightarrow \infty$ limit. This is constructed from \eqref{generallagr} considering only up to quadratic terms. For all the details omitted in the following about the superspace construction leading to the light-cone gauge fixed Lagrangian, we refer to \cite{Astolfi:2009qh}. Let us split up the Lagrangian in the bosonic and fermionic parts
\begin{equation}
\CL_2 = \CL_{2,B} + \CL_{2,F}
\end{equation}
and analyze these separately.

\subsection*{The bosonic sector}

The quadratic bosonic Lagrangian is
\begin{equation}
\label{CL2}
\CL_{2,B} = \frac{1}{2} \sum_{i=1}^4 ( \dot{u}_i^2 - {u_i'}^2 - c^2
u_i^2 ) + \frac{1}{16} \sum_{a=1}^2 ( \dot{x}_a^2 - {x_a'}^2 + 2 c
y_a \dot{x}_a + \dot{y}_a^2 - {y_a'}^2 )
\end{equation}
The momentum conjugate fields are defined by
\begin{equation}
\Pi_\mu = \frac{\partial \CL}{\partial \dot{x}^\mu}
\end{equation}
We get $\Pi_{x_a} = (\dot{x}_a + c y_a)/8$, $\Pi_{y_a} =
\dot{y}_a/8$ and $\Pi_{u_i} =\dot{u}_i$. By Legendre transforming the Lagrangian, the quadratic bosonic
Hamiltonian is obtained as
\begin{equation}
\label{CH2Bbis}
c \CH_{2,B} = \frac{1}{16} \sum_{a=1}^2 \Big[  p_{x_a} ^2 +
p_{y_a}^2  +   {x_a'}^2 + {y_a'}^2  \Big] + \frac{1}{2} \sum_{i=1}^4
\Big[ p_{u_i}^2 + {u_i'}^2 + c^2 u_i^2 \Big]
\end{equation}
where for convenience we defined the fields
\begin{equation}
p_{x_a}  \equiv 8 \Pi_{x_a} - c y_a \spa p_{y_a} \equiv 8 \Pi_{y_a}
\spa p_{u_i}  \equiv \Pi_{u_i}
\end{equation}
Notice that these fields are functions of the momenta and position
variables.

\subsection*{The fermionic sector}

It is useful to parameterize the fermionic directions in terms of a 16 components complex spinor, namely
\be
\label{defpsi}
\psi= \theta^1+i\Gamma_{049}\theta^2\,
\qquad
\psi^*= \theta^1-i\Gamma_{049}\theta^2\,
\ee
The gauge choice for the complex fermions is simply equivalent to $(\mathcal P_++\mathcal P_-)\psi=\psi$.
In the following we split up the spinor as
\begin{equation}
\psi = \psi_+ + \psi_-\,~~~{\rm with}~~~ \psi_\pm = \CP_\pm \psi
\label{psipm}
\end{equation}

The quadratic fermionic Lagrangian is, using the notation of \eqref{ABCdef}
\begin{equation}
\label{CL2Fb}
\CL_{2,F} = \frac{ic}{2} A_{+,\tau} + \frac{ic}{2}
\tilde{A}_{+,\sigma} + \frac{ic^2}{8} ( B_{+56}+B_{+78} ) -
\frac{i c^2}{4} C_{++}
\end{equation}
The physical degrees of freedom are singled out imposing light-cone gauge.
We get the following Lagrangian
\begin{eqnarray}
\label{finalCL2F}
\CL_{2,F} &=& \frac{ic}{2} \left[ \psi_+^* \dot{\psi_+} +2\psi_-^*\dot{\psi_-} - \frac{1}{2}\left(\psi_+ \psi_+' + \psi_+^* {\psi_+^*}'+2\psi_- \psi_-' +2\psi_-^* {\psi_-^*}'\right)    \right]  \nn \\ && + \frac{c^2}{4} \psi_+ \psi_+^* -c^2\psi_-\psi_-^*+ \frac{ic^2}{2} \psi_-\Gamma_{56}  \psi_-^*
\end{eqnarray}
%


\section{The Hamiltonian}
\label{app:Fermionsandmore}

In this Appendix we write down the explicit expressions for the terms in the expansion of the light-cone Hamiltonian~ \eqref{hgf} which have been computed in~\cite{Astolfi:2009qh}.

\subsection*{The purely bosonic Hamiltonian}

The bosonic plane-wave Hamiltonian is
\begin{equation}
\label{CH2B}
c \CH_{2,B} = \frac{1}{16} \sum_{a=1}^2 \Big[  p_{x_a} ^2 +
p_{y_a}^2  +   {x_a'}^2 + {y_a'}^2  \Big] + \frac{1}{2} \sum_{i=1}^4
\Big[ p_{u_i}^2 + {u_i'}^2 + c^2 u_i^2 \Big]
\end{equation}
The bosonic cubic and quartic Hamiltonians are
\begin{equation}
\label{CH3B}
\CH_{3,B} = \frac{u_4}{8c}  \Big[ p_{x_1}^2 + p_{y_1}^2 - p_{x_2}^2
- p_{y_2}^2  - {x_1'}^2 - {y_1'}^2 + {x_2'}^2 + {y_2'}^2 \Big]
\end{equation}
\begin{eqnarray}
\label{CH4B} && \CH_{4,B} = \frac{2}{c^3}(\sum_{i=1}^8 p_i
{X'}^i)^2-\frac{1}{2 c^3}\left(\sum_{i=1}^8( p_i^2+({X'}^i)^2)-c^2
\sum_{i=1}^3 u_i^2 + c^2u_4^2\right)^2 \cr && + c (\sum_{i=1}^3
u_i^2)^2+\frac{4}{3}cu_4^4+\frac{1}{c}\sum_{i,j=1}^3
u_i^2({u_j'}^2-p_j^2)+\frac{2}{c}u_4^2 \sum_{i=5}^8 p_i^2 \cr
&&+\frac{1}{12\sqrt{2}}(p_5y_1^3+p_7 y_2^3)+\frac{1}{2c}y_1^2
(p_5^2-{X_5'}^2)+\frac{1}{2c}y_2^2 (p_7^2-{X_7'}^2)
\end{eqnarray}
with
\begin{equation}
\begin{array}{c}
\ds p_{i=1...4} = (p_{u_1},p_{u_2},p_{u_3},p_{u_4}) \spa p_{i=5...8}
= \frac{\sqrt{2}}{4} (p_{x_1},p_{y_1},p_{x_2},p_{y_2}) \\
\ds {X'}^{i=1...4} = (u_1',u_2',u_3',u_4') \spa {X'}^{i=5...8} =
\frac{\sqrt{2}}{4} ( x_1',y_1',x_2',y_2' )
\end{array}
\end{equation}
%


\subsection*{The purely fermionic Hamiltonian}

In order to deal with the fermions it is useful to introduce the following notation
\begin{equation}
\label{ABCdef}
\begin{array}{c} \ds
A_{a,A} = \bar{\theta} \Gamma_a \partial_A \theta \spa
\tilde{A}_{a,A} = \bar{\theta} \Gamma_{11} \Gamma_a \partial_A
\theta \\[2mm] \ds B_{abc} = \bar{\theta} \Gamma_a \Gamma_{bc} \theta
\spa \tilde{B}_{abc} = \bar{\theta} \Gamma_{11} \Gamma_a \Gamma_{bc}
\theta
\\[2mm] \ds
C_{ab} = \bar{\theta}  \Gamma_a P \Gamma_{0123}
\Gamma_b  \theta \spa
\tilde{C}_{ab} =  \bar{\theta} \Gamma_{11}  \Gamma_a P
\Gamma_{0123} \Gamma_b  \theta
\end{array}
\end{equation}
\begin{equation}
\label{calABC}
\begin{array}{c} \ds
\CB_{abc;d} = \bar{\theta} \Gamma_{abc} ( \CP_+ + \frac{1}{2} \CP_- ) \Gamma^0 \Gamma_d \theta \spa \tilde{\CB}_{abc;d} = \bar{\theta} \Gamma_{11} \Gamma_{abc} ( \CP_+ + \frac{1}{2} \CP_- ) \Gamma^0 \Gamma_d \theta
\\[2mm] \ds
\CC_{ab;c} = \bar{\theta} \Gamma_a P \Gamma_{0123} \Gamma_b ( \CP_+ + \frac{1}{2} \CP_- ) \Gamma^0 \Gamma_c \theta \spa \tilde{\CC}_{ab;c} = \bar{\theta} \Gamma_{11} \Gamma_a P \Gamma_{0123} \Gamma_b ( \CP_+ + \frac{1}{2} \CP_- ) \Gamma^0 \Gamma_c \theta
\\[2mm] \ds
E_{ab} = \bar{\theta} \Gamma_a ( \CP_+ + \frac{1}{2} \CP_- ) \Gamma^0 \Gamma_b \theta \spa \tilde{E}_{ab} = \bar{\theta} \Gamma_{11} \Gamma_a ( \CP_+ + \frac{1}{2} \CP_- ) \Gamma^0 \Gamma_b \theta
\end{array}
\end{equation}
where $P$ is the projector defined in~\eqref{PP}, $\CP_{\pm}$ are defined in \eqref{pppm} and $a,b,\dots$ are the 10d tangent space indices.

The plane-wave fermionic Hamiltonian is
\begin{equation}
\label{CH2F}
\CH_{2,F} =
 \frac{i}{4 c^2} (c^2 \psi_+ \psi_+' -4 \rho_+ \rho_+' +2 c^2 \psi_- \psi_-' -2 \rho_- \rho_-')
 - \frac{i}{2} \psi_+ \rho_++i\psi_-\rho_-
 + \frac{1}{2} \psi_- \Gamma_{56} \rho_-
\end{equation}
where the conjugate momenta are
\be
\label{def_rho}
\rho \equiv \frac{\delta \CL_2}{\delta \dot{\psi}} = - \frac{ic}{2} ( 2 \CP_- + \CP_+ ) \psi^*
\ee
and
$\rho_{\pm}=\CP_{\pm}\rho$, cf. appendix \ref{app:plane_wave}.

The quartic purely fermionic Hamiltonian is
\begin{equation}
\label{pH4F}
\begin{array}{l} \ds
\CH_{4,F} = - \frac{i}{12} \Big( \bar{\theta} \Gamma_{11} \Gamma_+
\CM^2 \theta' + \bar{\theta} \Gamma_+ \CM^2 \Gamma_{11} \theta'
\Big)
- \frac{1}{2c} ( A_{+,\sigma}^2 - \tilde{A}_{+,\sigma}^2 )
\\[2mm] \ds
- \frac{1}{4} A_{+,\sigma} ( \tilde{C}_{+-} + \tilde{B}_{+56} + \tilde{B}_{+78} )
+ \frac{1}{4} \tilde{A}_{+,\sigma} ( C_{+-} - C_{++} + B_{+56} + B_{+78} )
\\[2mm] \ds
- \frac{c}{8} \sum_{i=1}^4 C_{+i}^2
- \frac{c}{32} \sum_{i=5}^8 \Big[ 2 C_{+i} - s_i B_{+4i} + \frac{1}{2} \sum_{j=5}^8 \epsilon_{ij} B_{+-j} \Big]^2
\end{array}
\end{equation}
where $\CM^2$ is defined in appendix \ref{app:plane_wave}.


\subsection*{The mixed bosonic-fermionic Hamiltonian}

The mixed cubic Hamiltonian is
\begin{eqnarray}
\label{finalH3BF}
\begin{array}{l} \ds
\CH_{3,BF} =  \frac{i}{2} \sum_{i=1}^8 (
 C_{+i}  p_i + \tilde{C}_{+i} {X'}^i )
- \frac{ic}{4} ( B_{+56} - B_{+78} ) u_4
 - \frac{ic}{4}  B_{+-4} u_4
\\[2mm] \ds
  - \frac{i }{4}
\sum_{i=5}^8 s_i ( B_{+4i} p_i + \tilde{B}_{+4i} {X'}^i )
- \frac{i}{8} \sum_{i,j=5}^8 \epsilon_{ij} ( B_{+-i} p_j +
\tilde{B}_{+-i} {X'}^j )
\end{array}
\end{eqnarray}
The mixed quartic Hamiltonian is
\begin{equation}
\label{finalH4BF}
\begin{array}{l} \ds
\CH_{4,BF} = \frac{i}{c^2}
\sum_{i=1}^8 \Big( p_i^2 + ({X'}^i)^2 \Big) \Big[ \tilde{A}_{+,\sigma}
 + \frac{c}{4} ( B_{+56} + B_{+78}  -  C_{++}  +  C_{+-} ) \Big]
- i \tilde{A}_{+,\sigma} \Big[ \sum_{i=1}^3 u_i^2 -u_4^2 \Big]
\\[2mm] \ds
+ \frac{2i}{c^2} \sum_{i=1}^8 p_i  {X'}^i \Big[  A_{+,\sigma} +
\frac{c}{4} ( \tilde{B}_{+56}+\tilde{B}_{+78}) + \frac{c}{4}
\tilde{C}_{+-}   \Big]
+ \frac{ic}{2} \sum_{i=1}^3 u_i^2 C_{++}
- \frac{ic}{4} \sum_{i=1}^4 u_i^2 ( B_{+56} + B_{+78})
\\[2mm] \ds
+ \frac{i}{2} u_4 \sum_{i=5}^8 s_i \Big[ C_{+i}  p_i  -
\tilde{C}_{+i} {X'}^i \Big]
- \frac{i}{c} \sum_{i,j=1}^8 \Big[  C_{ij} ({X'}^i {X'}^j - p_i  p_j
) + 2  \tilde{C}_{ij} {X'}^i p_j  \Big]
- i \sum_{i,j=1}^3 u_i' u_j \tilde{B}_{+ij}
\\[2mm] \ds
- \frac{i}{8} u_4 \sum_{i,j=5}^8 s_i \epsilon_{ij} ( 3 B_{+-i} p_j
+ \tilde{B}_{+-i} {X'}^j )
+ \frac{i}{4} ( B_{+56} p_{x_1} y_1 +  \tilde{B}_{+56} x_1' y_1 +  B_{+78} p_{x_2} y_2 +  \tilde{B}_{+78} x_2' y_2 )
\\[2mm] \ds
- \frac{i}{2} \sum_{i=1}^4 \sum_{j=1}^8 u_i \Big[ B_{-ij} p_j  - \tilde{B}_{-ij} {X'}^j \Big]
+ \frac{i}{2c} \sum_{i=1}^8 \sum_{j=5}^8 s_j \Big[ ( p_i  p_j  - {X'}^i {X'}^j ) B_{4 i j} + ( p_i  {X'}^j - {X'}^i p_j  ) \tilde{B}_{4 i j} \Big]
\\[2mm] \ds
- \frac{i}{2} \sum_{i=1}^3 \sum_{j=4}^8 u_i \Big[ B_{+ij} p_j -  \tilde{B}_{+ij} {X'}^j \Big]
- \frac{i}{4} u_4 \sum_{i=5}^8 ( B_{+4i} p_i  + 3 \tilde{B}_{+4i} {X'}^i )
+ \frac{i}{2} u_4 \sum_{i=1}^3 ( B_{+4i} p_i - \tilde{B}_{+4i} u_i' )
\\[2mm] \ds
- \frac{i}{4c} \sum_{i=1}^8 \sum_{j,k=5}^8 \epsilon_{jk} \Big[
(B_{+ij} - B_{-ij}) ( p_i  p_k  - {X'}^i {X'}^k )  +
(\tilde{B}_{+ij}-\tilde{B}_{-ij}) ( p_i  {X'}^k - {X'}^i p_k  )
\Big]
\\[2mm] \ds + \frac{i}{2 c^2  } \sum_{i,j=1}^8 (p_i p_j' + {X'}^i {X''}^j ) \tilde{E}_{ij} - \frac{i}{2 c^2  } \sum_{i,j=1}^8 ({X'}^i p_j' + p_i {X''}^j ) E_{ij} - \frac{3i}{4 c  } \sum_{i,j=1}^8 (p_i p_j - {X'}^i {X'}^j ) \CC_{i+;j}  %
\\[2mm] \ds
+ \frac{3i}{4 c  } \sum_{i,j=1}^8 ({X'}^i p_j - p_i {X'}^j ) \tilde{\CC}_{i+;j} - \frac{i}{4 c  } \sum_{i,j=1}^8 (p_i p_j + {X'}^i {X'}^j ) \CC_{+i;j} - \frac{i}{4 c  } \sum_{i,j=1}^8 ({X'}^i p_j + p_i {X'}^j ) \tilde{\CC}_{+i;j}
\\[2mm] \ds
+ \frac{i u_4}{2   } \sum_{i=1}^8 ( p_j \CB_{+-4;i} - {X'}^j \tilde{\CB}_{+-4;i} )
+  \frac{i}{2c  } \sum_{i=5}^8 \sum_{j=1}^8 s_i \Big[ (p_i p_j - {X'}^i {X'}^j) \CB_{+4i;j} + ({X'}^i p_j - p_i {X'}^j) \tilde{\CB}_{+4i;j}    \Big]
\\[2mm] \ds
+
\frac{i}{4c  } \sum_{i,j=5}^8 \sum_{k=1}^8 \epsilon_{ij} \Big[ (p_ip_k + {X'}^i {X'}^k)(\CB_{+-i;k} + E_{jk}) + ({X'}^i
p_k + p_i {X'}^k) (\tilde{\CB}_{+-i;k}-\tilde{E}_{jk}) \Big]
\end{array}
\end{equation}
In these expressions the fermionic coordinates are given in terms of the physical fermions and their conjugate momenta $\psi\,,\rho$
\begin{equation}
\label{thetheta}
\theta (\psi,\rho) = \frac{1}{2} ( \psi + E^{-1} \rho ) + \frac{\Gamma_{049}}{2i} ( \psi - E^{-1} \rho )
\end{equation}
where $E = - \frac{ic}{2} ( \CP_+ + 2\CP_- )$.

Eqs.~(\ref{CH2B}, \ref{CH3B},\ref{CH4B}, \ref{CH2F},\ref{finalH3BF}, \ref{finalH4BF}) are the starting point for the computation of the corrections to the energy of certain string states which is performed in this Paper.



\section{Gamma-matrix conventions}
\label{app:gamma}

Define the real $8 \times 8$ matrices $\gamma_1,...,\gamma_8$ as in
\cite{Callan:2003xr}. They obey
\begin{equation}
\begin{array}{c} \ds
\gamma_i \gamma_j^T + \gamma_j \gamma_i^T = \gamma_i^T \gamma_j +
\gamma_j^T \gamma_i = 2 \delta_{ij} I_8 \, , \, i,j=1,...,8
\\[4mm] \ds
\gamma_1 \gamma_2^T \gamma_3 \gamma_4^T \gamma_5 \gamma_6^T \gamma_7
\gamma_8^T = I_8 \spa \gamma_1^T \gamma_2 \gamma_3^T \gamma_4
\gamma_5^T \gamma_6 \gamma_7^T \gamma_8 = - I_8
\end{array}
\end{equation}
where $I_n$ is the $n\times n$ identity matrix. Define the $16
\times 16$ matrices $\hat{\gamma}_1,...,\hat{\gamma}_9$ by
\begin{equation}
\hat{\gamma}_i = \matrto{0}{\gamma_i}{\gamma_i^T}{0}\, , \,
i=1,...,8 \spa \hat{\gamma}_{9} = \matrto{I_8}{0}{0}{-I_8}
\end{equation}
The matrices $\hat{\gamma}_1,...,\hat{\gamma}_9$ are symmetric and
real and they obey
\begin{equation}
\{ \hat{\gamma}_i, \hat{\gamma}_j \} = 2 \delta_{ij} I_{16} \, , \,
i,j=1,...,9 \spa \hat{\gamma}_9 = \hat{\gamma}_1 \hat{\gamma}_2
\cdots \hat{\gamma}_8
\end{equation}
Define the $32 \times 32$ matrices
\begin{equation}
\Gamma_0  = \matrto{0}{-I_{16}}{I_{16}}{0} \spa \Gamma_i   =
\matrto{0}{\hat{\gamma}_i}{\hat{\gamma}_i}{0} \, , \, i=1,...,9 \spa
\Gamma_{11}  = \matrto{I_{16}}{0}{0}{-I_{16}}
\end{equation}
These matrices are real and obey
\begin{equation}
\{ \Gamma_a,\Gamma_b \} = 2 \eta_{ab} I_{32} \, , \,
i,j=0,1,...,9,11 \spa \Gamma_{11} = \Gamma^0 \Gamma^1 \cdots
\Gamma^9
\end{equation}
We define
\begin{equation}
\gamma_{i_1 \cdots i_{2k} } = \gamma_{[i_1} \gamma^T_{i_2}\cdots
\gamma^T_{i_{2k}]} \spa \gamma_{i_1 i_2 \cdots i_{2k+1} } =
\gamma^T_{[i_1} \gamma_{i_2} \cdots \gamma^T_{i_{2k+1}]} \spa i_l =
1,...,8
\end{equation}
\begin{equation}
\hat{\gamma}_{i_1 \cdots i_n } = \hat{\gamma}_{[i_1}
\hat{\gamma}_{i_2} \cdots \hat{\gamma}_{i_n]} \spa i_l = 1,...,9
\end{equation}
\begin{equation}
\Gamma_{i_1 i_2 \cdots i_n} = \Gamma_{[i_1} \Gamma_{i_2} \cdots
\Gamma_{i_n]} \spa i_l = 0,1,...,9,11
\end{equation}
%

\section{Small $c$ limit of $\CS(n)$ }
\label{app:smallc}

For completeness in this Appendix we report the small $c$ behavior of the sum $\CS(n)$ \eqref{csn}. We follow the procedure used in \cite{Grignani:2003cs} for a similar calculation in a different context.
By defining
\be
\CS_b(x)= \sum_{p=1}^\infty K_0(p x)
\ee
one can compute its Mellin transform, namely
\be
M_b(s)=\int_0^\infty dx x^{s-1} \CS_b(x) =\sum_{p=1}^\infty \int_0^\infty dx x^{s-1}  K_0(px)= \sum_{p=1}^\infty \int_0^\infty dx \int_0^\infty \frac{dt}{2t}\, x^{s-1}\, {e^{- t+\frac{x^2 p^2}{4t}}}~~~
\ee
where in the last step we have used a specific integral representation for the Modified Bessel Functions of the second kind  $K_0(x)$, cf. {\it e.g.}~\cite{gradsteyn}. After integrating we obtain
\be
M_b(s)=2^{s-2} \Gamma^2\left({s\over 2}\right)\sum_{p=1}^\infty  p^{-s} = 2^{s-2} \zeta(s)\Gamma^2\left({s\over 2}\right)
\ee
We can now perform the inverse Mellin transform of $M_b(x)$, namely
\be
\CS_b(x) = {1\over 2 \pi i} \int^{C+i\infty}_{C-i\infty} x^{-s} M_b(x) d s
\ee
The integral is well defined and to compute it we must close the contour
and use the residue theorem. For this purpose it is convenient to change the argument of
$\zeta(s)$ in the integrand according to~\cite{gradsteyn}
\be
\zeta(s)=\frac{\pi^{s-1/2}\Gamma\left(\frac{1-s}{2}\right)\zeta(1-s)}{\Gamma\left(\frac{s}{2}\right)}
\ee
so that
\be
\CS_b(x) = {1\over 2 \pi i} \int^{C+i\infty}_{C-i\infty} \left(\frac{2\pi}{x}\right)^{s} \frac{1}{4\sqrt{\pi}}\Gamma\left(\frac{1-s}{2}\right)\Gamma\left({s\over 2}\right)\zeta(1-s) d s
\ee
Closing the contour $C$ on the left we pick up the poles for $s=1,0,-2k,\dots,$ for $k=1, 2, 3,\dots$, which gives
\be
\CS_b(x)= {\pi \over 2 x}+{1\over 2} \left(\gamma-\log\left({4\pi\over x}\right)\right)+ \sum_{k=1}^\infty \frac{(-1)^k}{2\sqrt{\pi}\,k!} \, \left(\frac{x}{2\pi}\right)^{2 k}\, \,\Gamma\left(k+\frac{1}{2}\right)  {\zeta(2 k+1)}
\ee
We repeat the same technique for the other alternating sum
\be
\CS_f(x)= \sum_{p=1}^\infty (-1)^p K_0(p x)
\ee
and obtain a similar result
\be
\CS_f(x) = {1\over 2} \left(\gamma-\log \left({\pi\over x}\right)\right)+ \sum_{k=1}^\infty \frac{(-1)^k}{2\sqrt{\pi}\,k!}\, (2^{2 k+1}-1) \, \left(\frac{x}{2\pi}\right)^{2 k} \, \, \,\Gamma\left(k+\frac{1}{2}\right)  {\zeta(2 k+1)}
\ee
Hence the total sum $\CS(n)$ in Eq. \eqref{csn} reads
\be
\CS(n) &=& {8 n^2\over c\, R^2\, \omega_n} \left( \CS_f(\pi c ) -\CS_b (\pi c )\right) =
\\ \nn
&=& {8 n^2\over c\, R^2\, \omega_n}\left( \,- {1 \over 2 c} + \log 2 + \frac{1}{\sqrt{\pi}}\sum_{k=1}^\infty \frac{(-1)^k}{k!}\, (2^{2 k}-1) \, \left(\frac{c}{2}\right)^{2 k} \, \, \,\Gamma\left(k+\frac{1}{2}\right)  {\zeta(2 k+1)} \right)
\ee
Finally from the above expression one can easily read off the small $c$ leading behavior of $S(n)$
\begin{eqnarray}
\CS(n) &\simeq& -\frac{4 n}{c^2 R^2}+\frac{8 n \log (2)}{c R^2}+\frac{1}{2 n
   R^2}-\frac{c \left(3 n^2 \zeta (3)+\log (2)\right)}{n R^2}-\frac{3
   c^2}{32 \left(n^3 R^2\right)}+\CO\left(c^3\right)\cr&&
\end{eqnarray}

\end{appendix}


\addcontentsline{toc}{section}{References}



\providecommand{\href}[2]{#2}\begingroup\raggedright\endgroup

\end{document}